\documentclass[pra,aps,twocolumn,nopacs,superscriptaddress,nofootinbib]{revtex4}

\usepackage[T1]{fontenc}
\usepackage[latin1]{inputenc}
\usepackage{lmodern}
\usepackage{graphicx}
\usepackage{dcolumn}
\usepackage{bm}
\usepackage{amsmath}
\usepackage{amssymb}
\usepackage{pst-all}
\usepackage{psfrag}
\usepackage{epsfig}
\usepackage{units}
\usepackage{gensymb}
\usepackage{color}
\usepackage{array,multirow,graphicx}
\usepackage{float}

\def\ii{{\rm i}}  \def\ee{{\rm e}}
\def\Ree{{\rm Re}}  \def\Imm{{\rm Im}}
\newcommand{\abs}[1] {\mathopen{}\left|#1\right|\mathclose{}}
\newcommand{\sqabs}[1] {\mathopen{}\left|#1\right|^2\mathclose{}}
\newcommand{\ccpar}[1] {\mathopen{}\left(#1\right)\mathclose{}}
\newcommand{\sqpar}[1] {\mathopen{}\left[#1\right]\mathclose{}}
\newcommand{\clpar}[1] {\mathopen{}\left\{#1\right\}\mathclose{}}
\def\rb{{\bf r}}  \def\Rb{{\bf R}}    
\def\xx{\hat{\bf x}}  \def\yy{\hat{\bf y}}  \def\zz{\hat{\bf z}}
  \def\eh{\hat{\bf e}}
\def\RR{\hat{\bf R}}  
\def\kb{{\bf k}}  \def\kpar{k_\parallel}  \def\kparb{{\bf k}_\parallel}
    
\def\Eb{{\bf E}}    \def\Hb{{\bf H}}  
\def\pb{{\bf p}}  
\def\rp{r_{\rm p}}  \def\rs{r_{\rm s}}
    \def\EF{{E_{\rm F}}}
\def\lamp{\lambda_{\rm p}}      \def\kp{k_{\rm p}}  \def\wp{\omega_{\rm bulk}}
\def\epsilonm{\epsilon_{\rm m}}  \def\epsilonbar{\bar{\epsilon}}
\def\kparsp{k_{\rm p}}  \def\Sb{{\bf S}}
\def\Rp{{\mathcal R}_{\rm p}}  \def\aRp{|{\mathcal R}_{\rm p}|}
\newcommand*{\vcenteredhbox}[1]{\begin{tabular}{@{}c@{}}#1\end{tabular}}
\def\eh{\hat{\bf e}}
\def\vep{\epsilon}  \def\Ree{{\rm Re}}  \def\Imm{{\rm Im}}

\newcommand{\ind}[1] {{\mathrm{#1}}}
\newcommand{\vect}[1] {\boldsymbol{\mathbf{#1}}}
\newcommand{\vers}[1] {\boldsymbol{\mathbf{\hat{#1}}}}

\def\Omegared{{\color{red} \Omega}}
\definecolor{bluecol}{rgb}{0.0, 0.0, 0.0} 
\newcommand{\kz} {k_{z1}}
\newcommand{\Pl} {\mathcal{P}}
\newcommand{\Al} {\mathcal{A}}

\begin{document}

\title{Fundamental Limits to the Coupling between Light and 2D Polaritons \\ {\color{bluecol} by Small Scatterers}}

\author{Eduardo~J.~C.~Dias}
\affiliation{ICFO-Institut de Ci\'{e}ncies Fot\`{o}niques, The Barcelona Institute of Science and Technology, 08860 Castelldefels (Barcelona), Spain}
\author{F.~Javier~Garc\'{\i}a~de~Abajo}
\email{javier.garciadeabajo@nanophotonics.es}
\affiliation{ICFO-Institut de Ci\'{e}ncies Fot\`{o}niques, The Barcelona Institute of Science and Technology, 08860 Castelldefels (Barcelona), Spain}
\affiliation{ICREA-Instituci\'o Catalana de Recerca i Estudis Avan\c{c}ats, Passeig Llu\'{\i}s Companys 23, 08010 Barcelona, Spain}

\begin{abstract}
Polaritonic modes in two-dimensional van der Waals materials display short in-plane wavelengths compared with light in free space. As interesting as this may look from both fundamental and applied viewpoints, such large confinement is accompanied by poor in/out optical coupling, which severely limits the application of polaritons in practical devices. Here, we quantify the coupling strength between light and 2D polaritons in both homogeneous and anisotropic films using accurate rigorous analytical methods. In particular, we obtain universal expressions for the cross sections associated with photon-polariton coupling by point and line defects, as well as with polariton extinction and scattering processes. Additionally, we find closed-form constraints that limit the maximum possible values of these cross sections. Specifically, the maximum photon-to-plasmon conversion efficiency in graphene is $\sim10^{-6}$ and $\sim10^{-4}$ for point and line scatterers sitting at its surface, respectively, when the plasmon and photon energies are comparable in magnitude. We further show that resonant particles placed at an optimum distance from the film can boost light-to-polariton coupling to order unity. Our results bear fundamental interest for the development of 2D polaritonics and the design of applications based on these excitations.
\end{abstract}
%\date{\today}
%\pacs{78.67.-n,68.65.−k}
\maketitle
%68.65.−k Low-dimensional, mesoscopic, nanoscale and other related systems: structure and nonelectronic properties
%78.67.−n Optical properties of low-dimensional, mesoscopic, and nanoscale materials and structures

%\noindent \textbf{Keywords:}  2D polaritons, graphene plasmonics, optical coupling, fundamental limits, optical theorem

\section{INTRODUCTION}

The quest for optical excitations with increasing degree of spatial confinement has been recently fueled by the discovery of plasmons and other polaritonic modes sustained by atomically-thin two-dimensional (2D) van der Waals materials, such as graphene \cite{JBS09,JGH11,FRA12,paper196,YLZ12,LGA17,AND18} and hexagonal boron nitride (hBN) \cite{LDA18}. These excitations display in-plane wavelengths in the range of a few nanometers at infrared frequencies, with a $>100$-fold reduction in mode wavelength relative to free-space radiation \cite{paper235}. As a result of such high confinement, 2D surface polaritons (SPs) encompass a large spatial concentration of electromagnetic energy that becomes appealing for producing intense optical nonlinearities \cite{paper247,JC15,CYJ15} and strong interaction with quantum emitters \cite{paper176,NGG11,SRS18,paper318}. SPs further display high sensitivity to the environment that becomes useful to optically detect and identify small amounts of organic \cite{paper256,FAL16} and inorganic analytes \cite{HYZ16}. Additionally, because of the comparatively small number of atoms involved in a 2D nanostructure to support SPs, these excitations can be efficiently tuned by means of external stimuli, such as the potentials generated by electrical gates \cite{CPB11,JGH11,FRA12,paper196}, the exposure to magnetic fields \cite{YLL12}, and the introduction of optical heating \cite{paper235,NWG16,JKW16}, therefore holding great potential for applications in broad areas of optoelectronics \cite{GPN12}. These means of control also enable the exploration of new fundamental phenomena, which are further expanded by effectively creating materials with new electronic and optical properties through nanostructured gating \cite{FZW18} and layer stacking \cite{NFC12,NWW15}.

Despite the benefits of achieving a strong spatial confinement, the small wavelength of 2D SPs ($\lamp$) compared to free-space light ($\lambda_0$) implies a large in-plane momentum mismatch between them that needs to be compensated in order to enable the excitation of SPs through external illumination. A common approach to overcome this problem consists in using obstacles such as tips \cite{FAB11} and patterned nanostructures \cite{ANG14} that scatter light to produce induced evanescent fields with sufficient momentum to couple to propagating SP modes, effectively breaking the photon-polariton wavelength mismatch. In particular, point scatterers ({\it e.g.}, molecules, defects, and nanoparticles) are typically employed as basic elements to mediate such coupling, and besides their potential for the design of practical applications, they further provide a simple reference to quantify light-polariton coupling. Importantly, as we show below, point scatterers additionally permit us to formulate fundamental limits to the efficiency of the coupling process.

In this article, we reveal important fundamental limits to the coupling of light to 2D SPs assisted by scatterers such as point and line defects.  We present our results in the form of simple and rigorous analytical expressions, delivering an exact quantitative measure of photon-to/fom-polariton coupling and polariton-to-polariton scattering. To that end, we first introduce a universal characterization of 2D SPs and their associated optical fields in terms of a single parameter --the ratio of their wavelength to the film thickness--, regardless of the physical nature of the material's response. This allows us to determine the efficiency of different optical scattering channels involving incident light and polaritons, expressed in terms of universal light-polariton coupling strengths and fundamental limits to the scatterer polarizability. In particular, we find the maximum possible photon-to-polariton cross-section to be of the order of $\lamp^3/\lambda_0$ and $\lamp^2/\lambda_0$ for point and line scatterers, respectively. For graphene, the limits to in-coupling are quantified by the maximum possible ratio between the numbers of generated plasmons and incident photons $N_{\rm plasmon}/N_{\rm photon}\sim\alpha^3$ and $\sim \alpha^2$ for point and line scatterers, where $\alpha\approx1/137$ is the fine structure constant. Besides their fundamental interest, our results provide useful tools for the design of optical devices involving the in/out-coupling of propagating light and 2D SPs.

\begin{figure*}[htbp]
\begin{centering}
\includegraphics[width=0.90\textwidth]{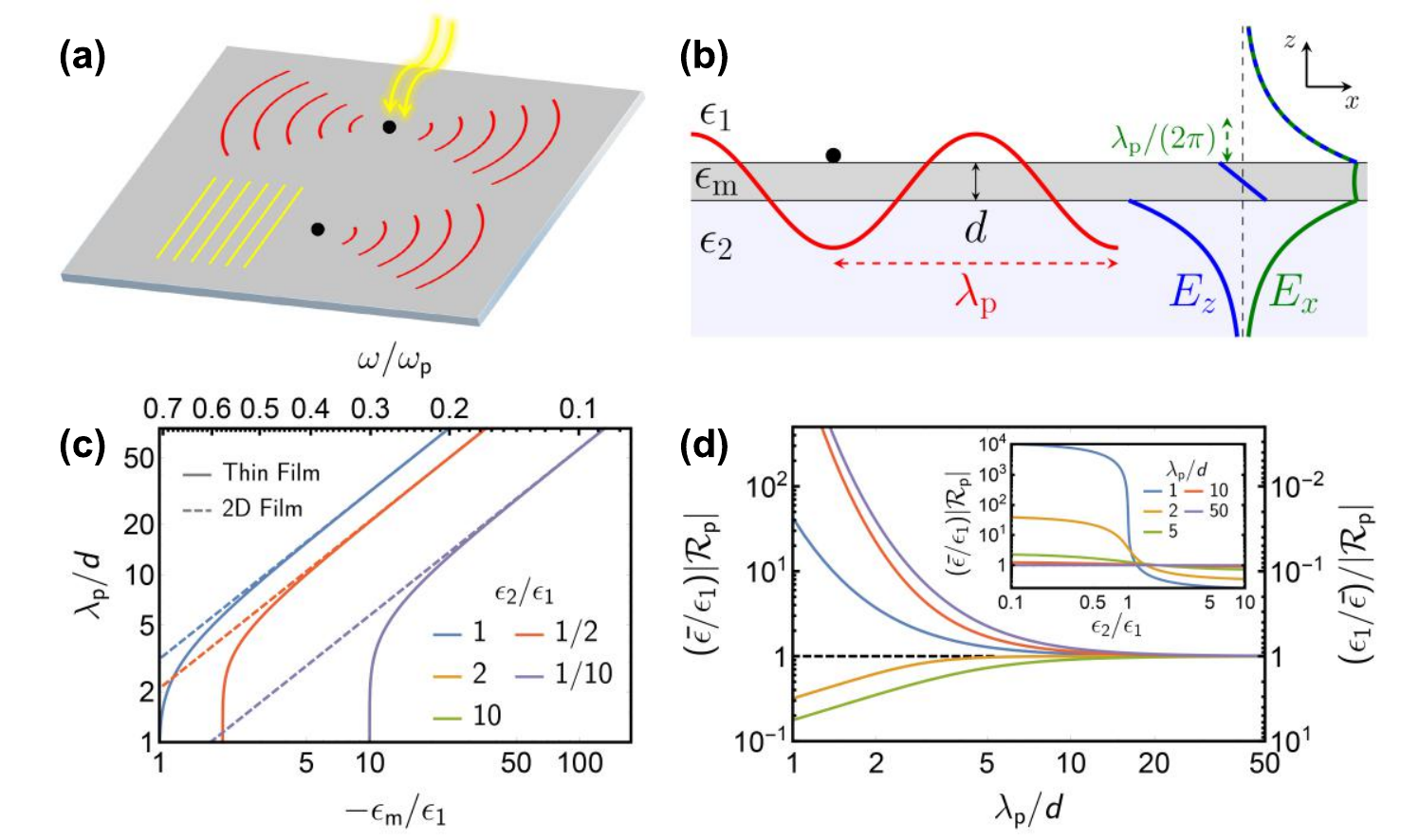}
\par\end{centering}
\caption{Polaritons in 2D materials and scattering by point defects. {\bf (a)} Schematic representation of the configurations under study. We consider a point scatterer (black dots) sitting on a thin film of finite thickness $d$. The scatterer is illuminated by either light or polaritonic waves (yellow), which produce an induced dipole that in turn gives rise to scattered polaritons (red). {\bf (b)} Basic properties of a 2D polariton: wavelength $\lamp$, field-amplitude decay distance $\sim\lamp/(2\pi)$ outside the film, and symmetry of in- and out-of-plane electric field components. {\bf (c)} Dispersion relation of the plasmon modes supported by a thin metallic layer (solid curves) calculated in the quasistatic limit and compared with the wavelength obtained in the zero-thickness limit (dashed curves). The latter is obtained by condensing the film permittivity $\epsilonm$ into a 2D conductivity (see main text). The upper horizontal axis gives the ratio $\omega/\wp$ corresponding to a Drude permittivity $\epsilonm(\omega)=1-\wp^2/\omega(\omega+\ii/\tau)$ for large relaxation time $\tau$ and $\epsilon_1=1$. {\bf (d)} Universal reflection-factor residue $\Rp$ (Eq.\ \eqref{rp}) for finite thickness under the conditions of (c) (see main text). The plots in (c,d) only depend on the permittivity contrast $\epsilon_1/\epsilon_2$, for which several values are considered (see legend in (c)).}
\label{Fig1}
\end{figure*}

\section{RESULTS AND DISCUSSION}

We first study the scattering of incident light or incident SPs by a point scatterer sitting at the surface of a thin film, as schematically sketched in Fig.\ \ref{Fig1}a. In particular, we consider the components of the scattered field that emerge from the scatterer as outgoing SPs. In what follows, we neglect retardation under the assumption that the SP modes are characterized by a large confinement ratio $\lambda_0/\lamp$ (see below Appendix\ \ref{sec:Retardation}). Then, the ratio $\lamp/d$ of SP wavelength to film thickness is only a function of the permittivities of the materials inside and outside the film, as illustrated in Fig.\ \ref{Fig1}c for isotropic metallic films. But first, before considering films of finite thickness in more detail, we investigate the useful $\lamp\ll d$ limit.

\subsection{Polaritons in Atomically Thin Layers}

Many of the properties of SPs in atomically thin layers (e.g, plasmons in graphene) can be accurately modeled in the zero-thickness limit, in which the material response is described {\it via} a frequency-dependent 2D conductivity $\sigma(\omega)$. In general, for a film material of permittivity $\epsilonm$ and thickness $d$, we can implicitly define a 2D conductivity $\sigma$ through the relation $\epsilonm=1+4\pi\ii\sigma/\omega d$. Figure\ \ref{Fig1}c shows that this approximation (dashed curves) yields an accurate prediction of the SP wavelength down to $\lamp\gtrsim10d$, a condition commonly fulfilled in few-atomic-layer films ($d\sim$\,nm) displaying SPs with $\lamp$ above a few tens of nanometers.

We thus focus first on a film of zero thickness characterized by a 2D conductivity $\sigma$ and placed in the $z=0$ plane. The electric field associated with a SP propagating along the in-plane direction $x$ then takes the form
\begin{align}
\Eb_{\rm p}(\rb)=E_0\; [\xx+\ii\;{\rm sign}\{z\}\zz]\; \ee^{\kparsp(\ii x-|z|)},
\label{Ep}
\end{align}
which clearly satisfies the Poisson equation $\nabla\cdot\Eb_{\rm p}=0$. Incidentally, we consider monochromatic fields of frequency $\omega$ with a temporal dependence given by $\Eb_{\rm p}(\rb,t)=2{\rm Re}\clpar{\Eb_{\rm p}(\rb)\ee^{-\ii\omega t}}$. This electric field describes oscillations along the film with wave vector $\kparsp$, corresponding to a SP wavelength $\lamp=2\pi/{\rm Re}\{\kparsp\}$, while it also implies a decay in intensity by a factor $1/\ee$ at a distance $\lamp/(4\pi)$ from the surface (Fig.\ \ref{Fig1}b). The parallel field component is continuous and symmetrically distributed on both sides of the film, whereas the perpendicular component is antisymmetric (Fig.\ \ref{Fig1}b) and presents a jump at the film plane $z=0$ in order to fulfill the boundary condition $\Delta(\epsilon E_z)=4\pi\rho^{\rm ind}$, where the induced surface charge $\rho^{\rm ind}$ is now determined from the induced surface current $j^{\rm ind}=\sigma E_x$ through the continuity equation $\rho^{\rm ind}=(-\ii/\omega)\partial_x j^{\rm ind}$; putting these expressions together, we readily derive the dispersion relation
\begin{align}
\kparsp=\frac{\ii\epsilonbar\,\omega}{2\pi\sigma}, \label{kparw}
\end{align}
where $\epsilonbar=(\epsilon_1+\epsilon_2)/2$ incorporates the effect of the dielectric environment on either side of the film (Fig.\ \ref{Fig1}b) {\color{bluecol} and we assume $\sigma$ to be isotropic}.

We remark that the use of the quasistatic limit is well justified upon examination of the 2D conductivity of atomic layers, which is generally described by the expression  $\sigma=\ii (e^2/\hbar)\omega_{\rm D}/(\omega+\ii/\tau-\omega_{\rm g})$, where the frequency $\omega_{\rm D}$ acts as a Drude weight ({\it e.g.}, $\omega_{\rm D}=\EF/(\pi\hbar)$ for graphene doped to a Fermi energy $\EF$, and $\omega_{\rm D}=\hbar nd/m^*$ for a Drude metal of thickness $d$, carrier density $n$, and effective mass $m^*$); the gap frequency $\omega_{\rm g}$ is associated with optical phonons or exitons but is zero in graphene and 2D metals; and $\tau$ is a phenomenological relaxation time. Indeed, upon insertion of this expression into the dispersion relation (Eq.\ \eqref{kparw}), we find $\lamp/\lambda_0=\alpha X\ll1$ for frequencies $\omega$ that make $X=\ccpar{2\pi/\epsilonbar}\,\omega_{\rm D}/\ccpar{\omega-\omega_{\rm g}}$ of order unity. In what follows, we further assume long relaxation time $\tau$, so that the SP propagation distance (for $1/\ee$ intensity decay) $L_{\rm p}=1/\left(2{\rm Im}\{\kparsp\}\right)=(\omega-\omega_{\rm g})\tau\lamp/(4\pi)$ is large compared with $\lamp$.

\begin{table*}[tb]
\centering
\begin{tabular}{|c|ccc|}
\hline &&& \\
{\color{blue} {\bf Geometry}} & {\color{blue} {\bf Scheme}} & {\color{blue} $\lamp/d$} & {\color{blue} $\Rp$}  \\ &&& \\
\hline &&& \\
\parbox[c]{35mm}{\color{blue} {\bf 2D Film} \\ (zero-thickness)}
& \vcenteredhbox{\includegraphics[width=25mm,angle=0,clip]{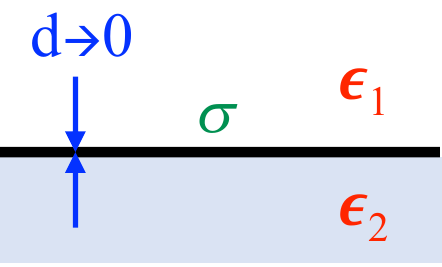}} & $4\pi^2\sigma/(\ii\epsilonbar\,\omega d)$  & $\epsilon_1/\epsilonbar$ \\ &&& \\ \hline &&& \\
{\color{blue} {\bf Thin Film}} & \vcenteredhbox{\includegraphics[width=25mm,angle=0,clip]{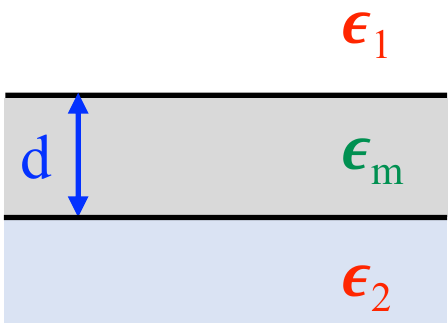}} & $4\pi\left(\log\sqpar{\frac{(\epsilon_1-\epsilonm)(\epsilon_2-\epsilonm)}{(\epsilon_1+\epsilonm)(\epsilon_2+\epsilonm)}}\right)^{-1}
$  & $\frac{\lamp}{\pi d}\frac{(-\epsilonm\epsilon_1)}{\epsilonm^2-\epsilon_1^2}$  \\ &&& \\ \hline &&& \\
{\color{blue} {\bf Anisotropic Film}} & \vcenteredhbox{\includegraphics[width=25mm,angle=0,clip]{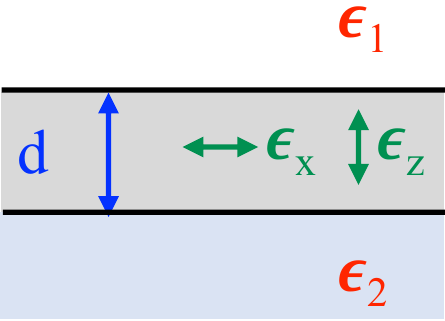}} & $4\pi\sqrt{\frac{\epsilon_x}{\epsilon_z}}\left(\log\sqpar{\frac{(\epsilon_1-\epsilonm)(\epsilon_2-\epsilonm)}{(\epsilon_1+\epsilonm)(\epsilon_2+\epsilonm)}}\right)^{-1}
$  & $\frac{\lamp}{\pi d}\frac{(-\epsilon_z\epsilon_1)}{\epsilon_x\epsilon_z-\epsilon_1^2}$ \\ &&& \\ \hline
\end{tabular}
\label{Table1}
\caption{Wavelength $\lamp$ and reflection residue $\Rp$ (Eq.\ \eqref{rp}) for SPs in different types of films (see Appendix\ \ref{sec:AppF}). For the 2D film we define $\epsilonbar=(\epsilon_1+\epsilon_2)/2$. For the anisotropic film we define $\epsilonm=\sqrt{\epsilon_x\epsilon_z}$ and take all square roots to yield positive imaginary parts.}
\end{table*}

\subsection{Polariton Coupling and Scattering}

The above results can be readily extended to films of finite thickness $d$ in the quasistatic limit ($\lamp\ll\lambda_0$), where $d$ provides a natural length scale in the system, and indeed, we find that light coupling can be expressed in terms of universal functions of the dimensionless ratio $\lamp/d$, which is in turn determined by the permittivities in the film and the surrounding media, as illustrated in Fig.\ \ref{Fig1}c for a metallic film. We concentrate here on SPs with p polarization ({\it e.g.}, plasmons), whose in-plane wave vector $\kparsp$ is identified from a pole in the corresponding Fresnel reflection coefficient,
\begin{align}
\rp\approx\Rp\,\frac{\kparsp}{\kpar-\kparsp},
\label{rp}
\end{align}
where $\Rp$ is a dimensionless residue that is equally determined by the dielectric response of the film and the surrounding materials. Specifically, in the zero-thickness limit, we have \cite{paper235} $\Rp=\epsilon_1/\epsilonbar$, while closed-form analytical expressions are equally obtained for homogeneous isotropic and anisotropic films (see Table\ \ref{Table1} and Appendix\ \ref{sec:AppF}).

In order to estimate the ability of small scatterers to couple light into SPs, we consider a point scatterer placed right on top of the film in the upper medium of permittivity $\epsilon_1$ (see Fig.\ \ref{Fig1}b). We describe the scatterer through an effective polarizability tensor $\alpha(\omega)$, which permits writing the dipole induced in the scatterer in response to an external electric field $\Eb^{\rm ext}$ as $\pb=\alpha\cdot\Eb^{\rm ext}$ . The field scattered by this dipole contains components of momentum overlapping the SP dispersion relation, which originate from the pole in Eq.\ \eqref{rp} ({\it i.e.}, $\kpar=\kparsp$), and indeed, are the only ones surviving at distant in-plane regions. Specifically, for large in-plane distance $R$ from the dipole ({\it i.e.}, $|\kparsp R|\gg1$) the field in medium 1 ($z>0$) reduces to (see Appendix\ \ref{sec:AppA})
\begin{align}
\Eb^{\rm scat}=E_\Rb^{\rm scat}\;(\RR+\ii\zz)\; \frac{\ee^{\kparsp(\ii R-z)}}{\sqrt{\kparsp R}},
\label{Eprad}
\end{align}
where
\[E_{\RR}^{\rm scat}=\sqrt{2\pi}\,\ee^{-\ii\pi/4}\,\frac{\Rp}{\epsilon_1}\kparsp^3\;(\ii\pb_\parallel\cdot\RR+p_z)\]
and $\pb_\parallel=(p_x,p_y)$. Equation\ \ref{Eprad} has a similar spatial dependence as Eq.\ \eqref{Ep}, with $\xx$ replaced by the unit radial vector $\RR$, and further incorporating a $1/\sqrt{R}$ decay that guarantees energy conservation for 2D circular waves. We seek to quantify the strength of coupling from incident light to SP modes by calculating the cross section defined by the ratio of power $P^{\rm scat}$ carried by the scattered surface modes to the incident light intensity $\sqrt{\epsilon_1}\,c|E^{\rm ext}|^2/(2\pi)$. The former can be evaluated by computing the Poynting vector for a circularly scattered wave (Eq.\ \eqref{Eprad}), which requires going beyond the quasistatic limit. Instead, we use a computationally simpler, yet rigorous alternative argument based on the decay rate of the induced dipole, which leads to (see Appendix\ \ref{sec:AppB})
\begin{align}
P^{\rm scat}=\frac{\aRp}{\epsilon_1}\frac{(2\pi)^4\omega}{\lamp^3}\ccpar{|\pb_\parallel|^2/2+|p_z|^2} \label{Pscat}
\end{align}
{\color{bluecol} and agrees with the computation based on the Poynting vector method (see detailed calculation in Appendix\ \ref{Poynting1}). This expression is dominated by surface polaritons arising from the pole of $\rp$ (see Appendix\ \ref{sec:AppA})}. Now, expressing the dipole moment in terms of the polarizability, we find that the point scatterer offers a photon-to-polariton (in-)coupling cross-section
\begin{widetext}
\begin{align}
\sigma^{\rm in-coup}_{\rm point}[{\rm area}]=\frac{\aRp}{\epsilon_1^{3/2}}\,\frac{(2\pi)^6}{\lamp^3\lambda_0} \;\times
\!\left\{ \begin{array}{ll}
\!\!A_-\cos^2\!\theta\,\sqabs{\alpha_x}/2+\!A_+\sin^2\!\theta\,\sqabs{\alpha_z},
& \text{(p-polarized light)}\\
\!\!A_0\sqabs{\alpha_x}/2,
& \text{(s-polarized light)}
\end{array} \right.
\label{sigmain}
\end{align}
\end{widetext}
where $\theta$ is the incidence angle relative to the film normal, we assume the polarizability to be diagonal in the $xyz$ frame with in- and out-of-plane components $\alpha_x=\alpha_y$ and $\alpha_z$, respectively, and the correction factors
\begin{subequations}
\begin{align}
A_\pm&=\sqabs{1\pm\rp}, \\
A_0&=\abs{1+\rs}^2
\end{align}
\label{AA}
\end{subequations}
are introduced to account for the interaction with incident light reflected by the film, as obtained by using the Fresnel coefficients $\rs$ and $\rp$ for s and p polarization (see Appendix\ \ref{sec:AppF}). The cross section in Eq.\ \eqref{sigmain} has units of area, so it can be interpreted as the portion of incident light plane wave that is effectively converted into SPs.

We can proceed in a similar way to calculate the cross section $\sigma^{\rm scat}_{\rm point}$ offered by the scatterer toward an impinging SP wave (see Eq.\ \eqref{Ep}) as the ratio $P^{\rm scat}/I_{\rm p}$, where $I_{\rm p}$ is the power flux carried by the SP plane wave ({\it i.e.}, the power per unit of transversal length along $y$). We compute this flux by comparing Eq.\ \eqref{Pscat} to the angular integral of the intensity associated with the scattered SP field (Eq.\ \eqref{Eprad}), which leads to (see Appendix\ \ref{sec:AppC})
\begin{align}
I_{\rm p}&=\frac{\epsilon_1}{\aRp}\,\frac{\omega\lamp^2}{(2\pi)^3} \sqabs{E_0}.
\label{Ip}
\end{align}
We now express the induced dipole as $\pb=E_0(\alpha_x\xx+\ii\alpha_z\zz)$ in Eq.\ \eqref{Pscat}, as determined by the incident plane wave (Eq.\ \eqref{Ep}), leading to the scattering cross-section
\begin{align}
\sigma^{\rm scat}_{\rm point}[{\rm length}]=\frac{\aRp^2}{\epsilon_1^2}\frac{(2\pi)^7}{\lamp^5} \ccpar{\sqabs{\alpha_x}/2+\sqabs{\alpha_z}},
\label{sigmascat}
\end{align}
which has units of length and represents the portion of the incident SP that is scattered into directions other than the incident one. Additionally, we find it useful to obtain the extinction cross-section for the incident SP
\begin{align}
\sigma^{\rm ext}_{\rm point}[{\rm length}]=\frac{\aRp}{\epsilon_1}\,\frac{16\pi^3}{\lamp^2} \mathrm{Im}\clpar{\alpha_x+\alpha_z},\label{sigmaext}
\end{align}
which results from the surface-integrated total ({\it i.e.}, incident+scattered) power flux along the forward-scattering direction (see Appendix\ \ref{sec:AppD}).

Incidentally, the above results can be trivially generalized to include a finite distance $z_0$ from the scatterer to the film by just multiplying the polarizability by $\ee^{-2\pi z_0/\lamp}$ and also correcting the coefficients $A_{\nu=\pm,0}$.

The effect of film thickness and material composition is captured both in $\Rp$ and in the frequency dependence of the $\lamp/d$ ratio. For a homogeneous film of nearly lossless metallic permittivity $\epsilonm$ ({\it i.e.}, $\Ree\clpar{\epsilonm}<0$ and $\Imm\clpar{\epsilonm}\ll\abs{\epsilonm}$), the SPs are plasmon polaritons characterized by (see Appendix\ \ref{sec:AppF} and Table\ \ref{Table1})
\begin{align}
\lamp/d&=4\pi/\log \sqpar{\frac{(\epsilon_1-\epsilonm)(\epsilon_2-\epsilonm)}{(\epsilon_1+\epsilonm)(\epsilon_2+\epsilonm)}} \label{lampd}
\end{align}
and
\begin{align}
\Rp&=\frac{\lamp}{\pi d}\frac{(-\epsilonm\epsilon_1)}{\epsilonm^2-\epsilon_1^2},
\label{Rphomo}
\end{align}
which we plot in Fig.\ \ref{Fig1}c,d for selected values of the permittivity contrast $\epsilon_2/\epsilon_1$. Reassuringly, we recover the zero-thickness limit ($d\rightarrow0$) by taking $\abs{\epsilonm}\rightarrow\infty$ while keeping $\epsilonm d$ finite, which leads to $\lamp/d\rightarrow-\pi\epsilonm/\epsilonbar$ from Eq.\ \eqref{lampd}, and consequently $\Rp\rightarrow\epsilon_1/\epsilonbar$ (see Fig.\ \ref{Fig1}d). Table\ \ref{Table1} also shows results for $\lamp$ and $\Rp$ corresponding to anisotropic films (see Appendix\ \ref{sec:AppF}), which directly apply to SPs in hyperbolic media \cite{LDA18}.

\begin{figure*}[htbp]
\begin{centering}
\includegraphics[width=0.80\textwidth]{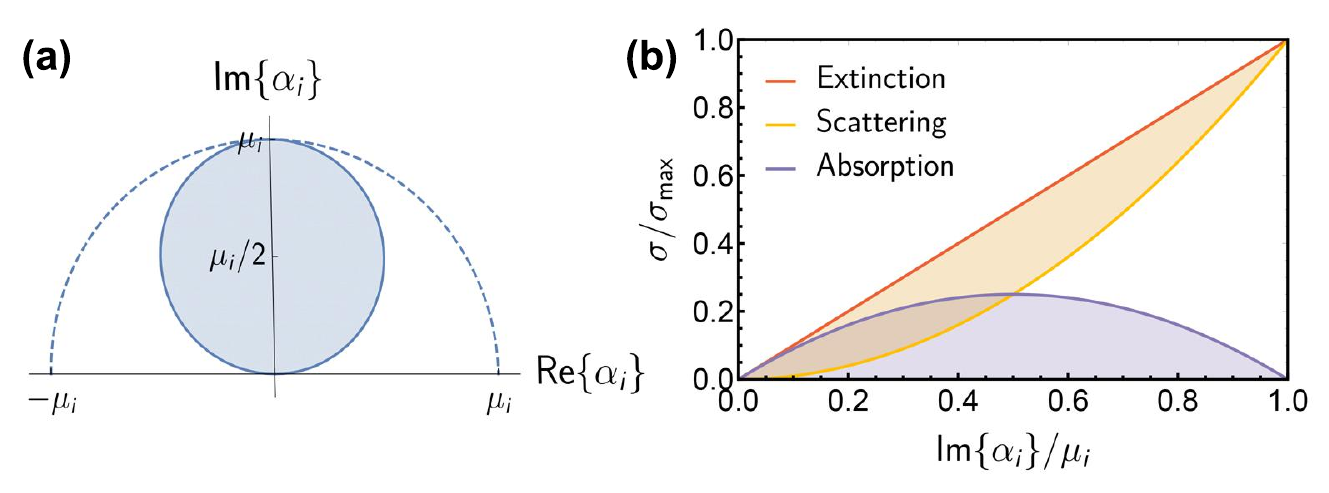}
\par\end{centering}
\caption{Fundamental limits to the polarizability and scattering cross-sections. {\bf (a)} The complex values of the polarizability are constrained to the shaded circle, whose diameter $\mu_i$ depends on the specific component: $\mu_z=\epsilon_1\lamp^3/(8\pi^4\aRp)$ for $\alpha_z$ and twice that value for $\alpha_x$. {\bf (b)} Variation of the extinction, scattering, and absorption cross-sections associated with the interaction of a polariton wave with a small scatterer as a function of the imaginary part of the scatterer polarizability. The horizontal axis is normalized to the maximum polarizability, whereas the vertical axis is normalized to the maximum extinction cross-section.}
\label{Fig2}
\end{figure*}

\subsection{Optical Theorem and Fundamental Limits for Polariton Scattering}
\label{sec:OT:SD}

Powerful constraints are imposed on the polarizatiblities of the scatterer from the fact that extinction is produced by both elastic scattering ({\it i.e.}, a change in the propagation direction of the incident polariton) and inelastic absorption, so we must have $\sigma^{\rm ext}\geq\sigma^{\rm scat}$. Using Eqs.\ \eqref{sigmascat} and \eqref{sigmaext}, this condition becomes $\mathrm{Im}\clpar{\alpha_x+\alpha_z}/(|\alpha_x|^2+2|\alpha_z|^2)\geq4\pi^4\aRp/(\epsilon_1\,\lamp^3)$, but in fact, separate conditions for in- and out-of-plane polarizability components can be extracted by considering two counter-propagating incident polariton waves with the same intensity, but with relative phases such that either the in- or the out-of-plane electric field component vanishes at the position of the scatterer. The induced dipole along the remaining non-vanishing component is increased by a factor of 2 relative to single-plane-wave incidence, leading to a 4-fold increase in the intensity of the elastically scattered component, also accompanied by a 2-fold increase in the extinction of each of the two incident waves. Putting these elements together, we find the conditions
\begin{align}
\mathrm{Im}\left\{\frac{-1}{\alpha_x}\right\}&\geq\frac{4\pi^4\aRp}{\epsilon_1\,\lamp^3}, \nonumber\\
\mathrm{Im}\left\{\frac{-1}{\alpha_z}\right\}&\geq\frac{8\pi^4\aRp}{\epsilon_1\,\lamp^3}, \nonumber
\end{align}
where $\Rp$ is defined in Eq.\ \eqref{rp} and explicitly calculated for a metallic film in Eq.\ \eqref{Rphomo}. These expressions, which constitute the optical theorem for 2D SPs, have the general form
\begin{align}
\mathrm{Im}\left\{-1/\alpha_i\right\}\geq1/\mu_i,
\label{optthe}
\end{align}
with $\mu_i$ just differing by a factor of 2 between in- ($i=x$, $\mu_x=\epsilon_1\lamp^3/(4\pi^4\aRp)$) and out-of-plane ($i=z$, $\mu_z=\epsilon_1\lamp^3/(8\pi^4\aRp)$) polarization components. Equation\ \eqref{optthe} can be recast as $\left(\mathrm{Re}\left\{\alpha_i\right\}\right)^2+\left(\mathrm{Im}\left\{\alpha_i\right\}-\mu_i/2\right)^2\leq(\mu_i/2)^2$, which clearly reveals that the possible values of $\alpha_i$ lie within a circle or radius $\mu_i/2$ centered around $\alpha_i=\ii\mu_i/2$ in the complex plane, as shown in Fig.\ \ref{Fig2}a. We conclude that the modulus of the polarizability and its imaginary part are both simultaneously maximized if ${\rm Re}\clpar{\alpha_i}=0$ and ${\rm Im}\clpar{\alpha_i}=\mu_i$, corresponding to the top point of the circle represented in that figure. Direct application of these results to a scatterer with polarization either parallel or perpendicular with respect to the film leads to the relation
\begin{align}
{\rm max}\clpar{|\alpha_z|}=\frac{1}{2}{\rm max}\clpar{|\alpha_x|}=\frac{\epsilon_1\lamp^3}{8\pi^4\aRp}.
\label{OTfinal}
\end{align}
The maximum achievable polarizabilty is thus larger for the in-plane component (${\rm max}\{|\alpha_x|\}>{\rm max}\{|\alpha_z|\}$), a result that we relate to its poorer coupling to SPs; indeed, a Lorentzian excitation in the scatterer decays more slowly to SPs for in-plane polarization, so it can have longer lifetime $\tau$, and consequently, also larger on-resonance polarizability $\propto\tau$. Additionally, for scatterers with the strongest possible polarizability $\alpha_i=\ii\mu_i$, we find
\begin{align}
{\rm max}\clpar{\sigma^{\rm scat}_{\rm point}}&={\rm max}\clpar{\sigma^{\rm ext}_{\rm point}}\nonumber\\&=\frac{2\lamp}{\pi}\times\left\{ \begin{array}{l l}
2, & \quad \text{for $x$ polarization}\\
1, & \quad \text{for $z$ polarization}\\
\end{array} \right.
%\nonumber
\end{align} 
depending on the orientation of the scatterer polarization. Remarkably, this result is universal, independent of thickness and dielectric properties of the film, and valid even when retardation is taken into consideration (see Appendix\ \ref{InfluenceofSize}).

The difference between extinction and scattering represents absorption by the scatterer. For polarization only along one direction $i=x$ or $z$, the absorption cross-section $\sigma^{\rm abs}_{\rm point}=\sigma^{\rm ext}_{\rm point}-\sigma^{\rm scat}_{\rm point}$ is thus the difference of two terms linear in ${\rm Im}\clpar{\alpha_i}$ and $\sqabs{\alpha_i}$, respectively, so it admits a maximum as a function of the polarizability determined by $\partial\sigma^{\rm abs}_{\rm point}/\partial\alpha_i=0$. This condition is trivially satisfied for ${\rm Re}\clpar{\alpha_i}=0$ and ${\rm Im}\clpar{\alpha_i}=\mu_i/2$, which leads to ${\rm max}\clpar{\sigma^{\rm abs}_{\rm point}}={\rm max}\clpar{\sigma^{\rm ext}_{\rm point}}/4$. Interestingly, this value of the polarizability produces the same cross-section for scattering and absorption, {\color{bluecol} satisfying what is known as the critical-coupling condition.}

We find it interesting to present the extinction, scattering, and absorption cross-sections normalized to the maximum extinction $\sigma_{\rm max}\equiv{\rm max}\clpar{\sigma^{\rm ext}_{\rm point}}$ as functions of the normalized imaginary part of the polarizability $\zeta={\rm Im}\clpar{\alpha_i}/\mu_i$, as shown in Fig.\ \ref{Fig2}b. Obviously, extinction corresponds to the straight line $\sigma^{\rm ext}_{\rm point}/\sigma_{\rm max}=\zeta$. Also, for each value of $\zeta$ the minimum scattering and correspondingly the maximum absorption are obtained when ${\rm Re}\clpar{\alpha_i}=0$, leading to the limiting values $\sigma^{\rm scat}_{\rm point}/\sigma_{\rm max}=\zeta^2$ and $\sigma^{\rm abs}_{\rm point}/\sigma_{\rm max}=\zeta-\zeta^2$, respectively (see solid curves in Fig.\ \ref{Fig2}b). However, ${\rm Re}\clpar{\alpha_i}$ can take nonzero values, only limited by the condition that $\alpha_i$ lies within the circle of Fig.\ \ref{Fig2}a, so by sweeping this parameter we obtain the colored regions presented in Fig.\ \ref{Fig2}b for the possible ranges of scattering and absorption cross-sections.

Incidentally, we have verified that Fig.\ \ref{Fig2}b remains valid even when including retardation (see Appendix\ \ref{InfluenceofSize}), so it can be regarded as a universal result, provided the extinction and scattering cross-sections are proportional to ${\rm Im}\clpar{\alpha_i}$ and $\sqabs{\alpha_i}$, respectively, and the particle polarizability satisfies an optical-theorem constrain of the form given by Eq.\ \eqref{optthe}. This is for example the case of a particle in free space, for which the optical theorem is equally given by Eq.\ \eqref{optthe} with $\mu_i=3\lambda_0^3/(16\pi^3)$ \cite{J99}.

\subsection{Limits to Light-Polariton Coupling}

\begin{figure*}[htbp]
\begin{centering}
\includegraphics[width=0.80\textwidth]{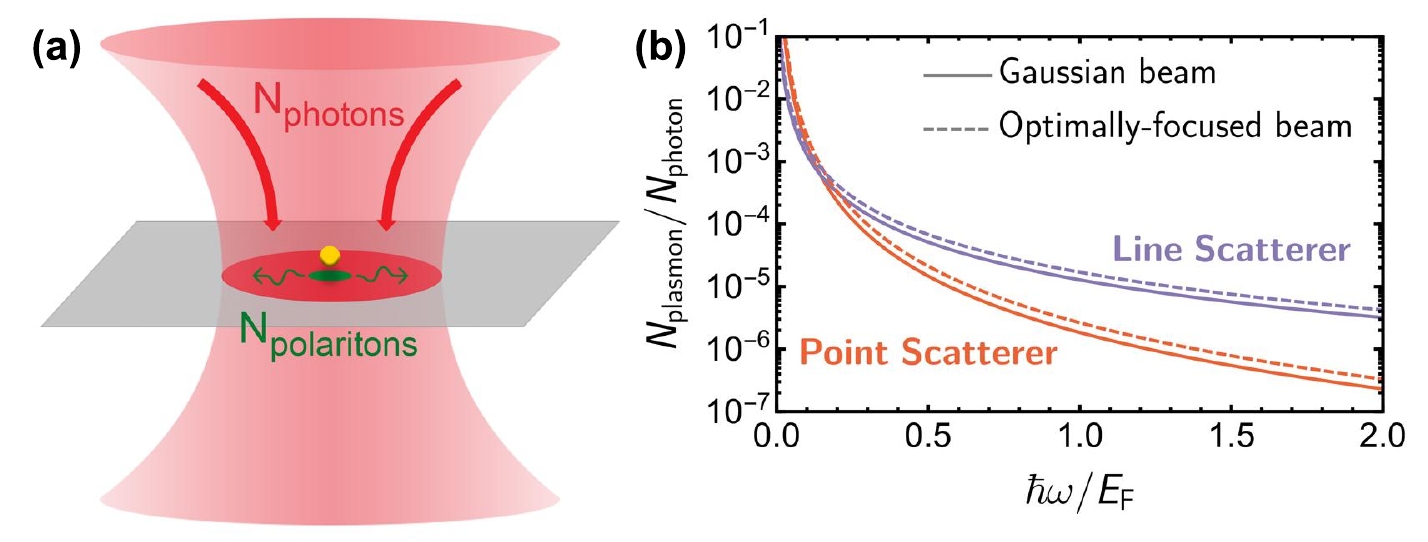}
\par\end{centering}
\caption{Maximum photon-to-polariton coupling efficiency. {\bf (a)} Schematic illustration of a tightly-focused external light beam interacting with a point defect. The red circle represents the beam spot (area $\sim\lambda_0^2$), while the green circle is the coupling cross-section ($\sim\lamp^3/\lambda_0$). {\bf (b)} Maximum number of plasmons generated per incident photon $N_{\mathrm{plasmon}}/N_{\mathrm{photon}}$ for coupling to point (orange curves) or line (purple curves) defects in graphene as a function of photon energy normalized to the doping Fermi energy $\EF$. We consider beams with either Gaussian (solid curves) or optimally-focused (dashed curves) profiles.} \label{Fig3}
\end{figure*}

Inserting the maximum values of the polarizability obtained in the previous section (Eq.\ \eqref{OTfinal}) into Eq.\ \eqref{sigmain}, we find the maximum normal-incidence photon-to-polariton coupling cross-section
\begin{align}
{\rm max}\clpar{\sigma^{\rm in-coup}_{\rm point}} = \frac{2\sqrt{\epsilon_1}A_0}{\pi^2\aRp}\,\frac{\lamp^3}{\lambda_0}.
\label{maxsigmain}
\end{align}
This result imposes a severe reduction in the possible coupling of incident photons to polaritons when we compare it to the minimum focal spot in tightly-focused light beams (see Fig.\ \ref{Fig3}a).

We find it useful to estimate the maximum coupling efficiency by computing the ratio of polaritons produced per incident photon, $N_{\rm polariton}/N_{\rm photon}$. This quantity must be equal to the ratio of SP-scattered to incident powers $P^{\rm scat}/P^{\rm beam}$. However, the former is proportional to the light intensity at the position of the scatterer, which obviously depends on the angular profile of the light beam. We consider two relevant configurations for light focusing: a standard diffraction-limited Gaussian beam \cite{NH06} and an optimally-focused beam that maximizes the intensity at the spot center, for which we find a ratio of focal-spot intensity to beam power given by $|E_x|^2/P^{\rm beam}=g\,k^2\sqrt{\epsilon_1}/c$ with $g=0.4646$ and $2/3$, respectively (see Appendix\ \ref{sec:AppG}), where we ignore reflection at the film surface for simplicity. Now, we evaluate $P^{\rm scat}$ using Eq.\ \eqref{Pscat}, where we consider an incident beam with electric field along $x$ in the focal spot ({\it i.e.}, $|p_x|=\sqrt{A_0}\,|\alpha_x E_x|$ and $p_y=p_z=0$). Using the maximum in-plane polarizability (Eq.\ \eqref{OTfinal}), we find for the normal-incidence coupling efficiency the fundamental limit
\begin{align}
\frac{N_{\rm polariton}}{N_{\rm photon}}=
\frac{4g\epsilon_1^{3/2}A_0}{\pi\aRp}\left(\frac{\lamp}{\lambda_0}\right)^3. \label{NN1}
\end{align}
Because of the wide interest driven by plasmon polaritons in graphene, we now apply Eq.\ \eqref{NN1} to these excitations, whose wavelength depends on the doping Fermi energy $\EF$ as $\lamp/\lambda_0=(2\alpha/\epsilonbar)\EF/\hbar\omega$ \cite{paper235}. In this material, we can safely apply the zero-thickness limit ($\Rp=\epsilon_1/\epsilonbar$) and neglect beam reflection ($A_0=\sqabs{1+\rs}=1$) to find
\begin{align}
\left.\frac{N_{\rm plasmon}}{N_{\rm photon}}\right|_{\rm graphene}=\alpha^3\,
\frac{32g\sqrt{\epsilon_1}}{\pi\epsilonbar^2}\left(\frac{\EF}{\hbar\omega}\right)^3, \nonumber
\end{align}
which becomes of the order of $\alpha^3\sim10^{-6}$ for $\hbar\omega\sim\EF$. This result is plotted in Fig.\ \ref{Fig3}b for both Gaussian and optimally-focused beams, showing a dramatic boost in efficiency as the photon energy is reduced relative to $\EF$, reaching a value $N_{\rm plasmon}/N_{\rm photon}\approx0.26\%$ for self-standing graphene ($\epsilon_1=\epsilon_2=1$) at $\hbar\omega=0.1\,\EF$. 

The polariton-to-photon (out-)coupling efficiency is also an important magnitude that we can work out from the above results using the reciprocity theorem, or alternatively, by integrating the far-field Poynting vector generated by the scatterer upon SP irradiation ({\it i.e.}, $\sqrt{\epsilon_1}(2\pi)^4c|\pb|^2/(3\lambda_0^4)$ for upper-hemisphere emission). Dividing this result by the SP power flux $I_{\rm p}$ (Eq.\ \eqref{Ip}), noticing that $\pb=E_0(\alpha_x\xx+\ii\alpha_z\zz)$ is the dipole produced upon SP plane wave irradiation, and neglecting again surface reflection for simplicity, we find an associated cross-section \[\sigma^{\rm out-coup}_{\rm point}[{\rm length}]=\frac{\aRp}{3\sqrt{\epsilon_1}}\,\frac{(2\pi)^6}{\lamp^2\lambda_0^3}
\ccpar{\sqabs{\alpha_x}+\sqabs{\alpha_z}},\] and from here, using Eq.\ \eqref{OTfinal}, a maximum out-coupling efficiency
\begin{align}
{\rm max}\clpar{\sigma^{\rm out-coup}_{\rm point}}= \frac{4\epsilon_1^{3/2}}{3\pi^2\aRp}\,\frac{\lamp^4}{\lambda_0^3}
\nonumber
\end{align}
is obtained for a scatterer with in-plane polarization and $1/4$ times this value for out-of-plane orientation.

\subsection{Coupling through a Line Defect}

Line defects offer a way of generating polariton plane waves and constitute interesting elements to control polariton propagation. Scattering by line defects in graphene has been considered following semi-analytical methods \cite{RK15,FRK16}. Here, we can obtain limits to polariton coupling and scattering by line defects by representing them as a chain of closely spaced dipoles, for which we can use the results obtained in the preceding paragraphs. We thus define the polarizability and dipole densities $\mathcal{A}=\alpha/L$ and $\vec{\mathcal{P}}=\pb/L$ for a line defect extending along $y$ by normalizing to the defect length $L$. Assuming incident light or SP fields under normal incidence with respect to $y$ for the sake of concreteness, the scattered field is easily obtained by integrating Eq.\ \eqref{Eprad} along $y$ in the $|\kparsp x|\gg1$ limit, which produces an electric field like Eq.\ \eqref{Ep} with $E_0$ replaced by
\begin{align}
E^{\rm scat}_{\rm line}=\frac{8\pi^3\Rp}{\epsilon_1\lamp^2}\,(\ii\mathcal{P}_x+\mathcal{P}_z)
\label{Escatline}
\end{align}
(see Appendix\ \ref{sec:AppE}). Obviously, polarization along $y$ does not couple to SPs propagating along $x$.

For SP incidence (see Eq.\ \eqref{Ep}), the dipoles take the form $\mathcal{P}_x=\mathcal{A}_xE_0$ and $\mathcal{P}_z=\ii\mathcal{A}_zE_0$, which upon insertion into the scattered field permit us to obtain the reflection and transmission coefficients of the line scatterer, $r=E^{\rm scat}_{\rm line}/E_0=\ii\,[8\pi^3\Rp/(\epsilon_1\lamp^2)]\,\ccpar{\mathcal{A}_x+\mathcal{A}_z}$ and $t=1+r$, respectively. We can now obtain limits to the line polarizability by imposing the condition of non-negative absorption $1-\sqabs{t}-\sqabs{r}\geq0$, which results in $\mathrm{Im}\clpar{\mathcal{A}_x+\mathcal{A}_z}/(|\mathcal{A}_x+\mathcal{A}_z|^2)\geq8\pi^3\aRp/(\epsilon_1\lamp^2)$. Arguing again that we can superimpose two counter-propagating SPs that cancel either the $x$ or $z$ field component, we obtain two individual conditions for the polarizabilities $\mathcal{A}_i$ expressed by Eq.\ \eqref{optthe} with
\begin{align}
{\rm max}\clpar{|\mathcal{A}_i|}=\mu_i=\epsilon_1\lamp^2/(8\pi^3\aRp).
\label{maxAline}
\end{align}
We note that this condition is the same for both $i=x$ and $i=z$.

Incidentally, a maximum possible absorption of $50\%$ is achieved for $r=-1/2$, which implies $\mathrm{Re}\clpar{\mathcal{A}_x+\mathcal{A}_z}=0$ and $\mathrm{Im}\clpar{\mathcal{A}_x+\mathcal{A}_z}=\mu_i/2$. This condition is compatible with the optical theorem just derived for the line scatterer ({\it i.e.}, $\mathrm{Im}\clpar{\mathcal{A}_i}\leq\mu_i$). Additionally, we find $|r|\leq1$, with the equality being reached if the polarizability takes the maximum possible value $\mathcal{A}_i=\ii\mu_i$.

The photon-to-polariton coupling cross-section is now given as the ratio of the SP scattered power $P^{\rm scat}$ to the incident light intensity. Using Eq.\ \eqref{Ip} with $E_0$ substituted by $E^{\rm scat}_{\rm line}$, we find $P^{\rm scat}=2\times[L\epsilon_1\omega\lamp^2/(8\pi^3\aRp)]\, \sqabs{E^{\rm scat}_{\rm line}}$, where the leading factor of 2 accounts for the fact that the line scatters SPs toward either side of it. Considering normal incidence for simplicity, normalizing to the line length $L$, and noting that $|E^{\rm scat}_{\rm line}|=|E^{\rm ext}|$ when $E^{\rm scat}_{\rm line}$ (Eq.\ \eqref{Escatline}) is evaluated for the maximum possible polarizability (Eq.\ \eqref{maxAline}), we find
\begin{align}
{\rm max}\clpar{\sigma^{\rm in-coup}_{\rm line}}[{\rm length}]=\frac{\sqrt{\epsilon_1}}{\pi\aRp}\,\frac{\lamp^2}{\lambda_0}.
\nonumber
\end{align}
for the maximum possible in-coupling cross-section.

We can also convert the above result into a photon-to-polariton conversion efficiency by invoking the relation $|E_x|^2L/P^{\rm beam}=g\,k/c$ between the electric field intensity at the focus and the beam power per unit length $L$ for Gaussian ($g=0.1887$) and optimally-focused ($g=1/4$) line beams (see Appendix\ \ref{sec:AppG}) using the same procedure as for point scatterers. We find $N_{\rm polariton}/N_{\rm photon}=[(g\epsilon_1/(\pi\aRp)]\,(\lamp/\lambda_0)^2$. For graphene plasmons, this ratio becomes
\begin{align}
\left.\frac{N_{\rm plasmon}}{N_{\rm photon}}\right|_{\rm graphene}=\alpha^2\,
\frac{4g}{\pi\epsilonbar}\left(\frac{\EF}{\hbar\omega}\right)^2, \nonumber
\end{align}
which is of the order of $\alpha^2\sim10^{-4}$ for $\hbar\omega\sim\EF$, but can be boosted at low frequencies as shown in Fig.\ \ref{Fig3}b. Comparing the value $N_{\rm plasmon}/N_{\rm photon}\approx0.16\%$ obtained from this expression for $\hbar\omega=0.1\,\EF$ in self-standing graphene with 0.26\% for point scatterers (see above), line scatterers are less efficient per photon at the energy under consideration and become even comparatively worse as the ratio $\hbar\omega/\EF$ is reduced.

Out-coupling from polaritons to photons is also limited by the optical theorem, leading to a maximum conversion fraction $N_{\rm photon}/N_{\rm polariton}=[\epsilon_1/(4\aRp)]\,\left(\lamp/\lambda_0\right)^2$, as obtained by normalizing the upward power flux radiated by the line of induced dipoles $(2\pi^3\omega/\lambda_0^2)\,|{\bf\mathcal{P}}|^2$ (see Appendix\ \ref{sec:AppE}) to the SP plane wave power flux (Eq.\ \eqref{Ip}), assuming the maximum possible polarizabilities $|\mathcal{A}_i|=\mu_i$. We note that this fraction has the same order of magnitude and wavelength scaling $\sim\left(\lamp/\lambda_0\right)^2$ as the in-coupling fraction for a tightly-focused light beam.

\subsection{Coupling through a Linear Edge}

The linear edge of a film constitutes a special case of line scatterer. In the zero-thickness limit for a self-sustained semi-infinite film, using the notation of Eq.\ \eqref{Ep}, the polariton amplitude has been shown to satisfy the relation $E_0=\sqrt{2}E^{\rm ext}$ relative to a normally incident light field $E^{\rm ext}$ \cite{ZFY14}. Converting this relation to a photon-to-polariton cross-section (with units of length, as it is normalized to the edge length $L$) and using Eq.\ \eqref{Ip}, we find $\sigma^{\rm in-coup}_{\rm edge}=\lamp^2/(\pi\lambda_0)$, which coincides with ${\rm max}\clpar{\sigma^{\rm in-coup}_{\rm line}}$ (with parameter $\epsilon_1=\epsilon_2=1$ and $\Rp=1$ appropriate for a self-standing thin film), indicating that the edge is already attaining the maximum possible coupling efficiency, with all of the generated polaritons obviously directed only along one direction away from the edge.

\subsection{Optimization of Light-SP Coupling through Scatterer-Film Separation}

An interesting observation comes from the fact that the point scattering efficiency depends on the separation $z_0$ between the scatterer and the film surface through an exponential factor $\ee^{-\kparsp z_0}\approx\ee^{-2\pi z_0/\lamp}$ in the SP field. {\color{bluecol} In particular, the scatterer dipole induced by an incident SP plane wave then becomes $\pb=\ee^{-2\pi z_0/\lamp}\,E_0(\alpha_x\xx+\ii\alpha_z\zz)$. Repeating the analysis of the preceding sections with this factor in mind, we find the proportionalities $\sigma^{\rm ext}_{\rm point}\propto\ee^{-4\pi z_0/\lamp}\,{\rm Im}\{\alpha\}$ and $\sigma^{\rm scat}_{\rm point}\propto\ee^{-8\pi z_0/\lamp}\,|\alpha|^2$, and consequently, the optical theorem arising from $\sigma^{\rm ext}_{\rm point}\ge\sigma^{\rm scat}_{\rm point}$ leads to $\max\clpar{\alpha}\propto\ee^{4\pi z_0/\lamp}$, which produces maximum possible extinction and scattering cross-sections independent of $z_0$, while $\max\clpar{\sigma^{\rm in-coup}_{\rm point}}\propto\ee^{-4\pi z_0/\lamp}\,|\alpha|^2\propto\ee^{4\pi z_0/\lamp}$ increases exponentially with the separation $z_0$. Consequently, a finite $z_0$ allows us to enhance the light-polariton coupling efficiency without affecting polariton-polariton processes.

This result for the maximum possible in-coupling cross-section is however conditioned to finding particles that can actually reach the predicted limit of $\max\clpar{\alpha}$. As we show below, an optimum distance can be found when considering a realistic model for the scatterer polarizability. In practice, actual scatterers ({\it e.g.}, nanoparticles or molecules placed close to the film, or even protrusions and corrugations) present additional intrinsic loss channels, which limit the exponential increase with $z_0$ just predicted for $\sigma^{\rm in-coup}_{\rm point}$. In order to illustrate this point, we consider a particle hosting a spectrally-isolated resonance. As shown in Appendix\ \ref{sec:AppH}, the particle polarizability including its interaction with the film can be approximated as
\begin{align}
\alpha=\frac{p_0^2/\hbar}{\tilde\omega_0-\omega-\ii(\gamma_{\rm in}+\gamma_{\rm homo}+\gamma_{\rm film})/2},
\label{alphafull}
\end{align}
where $\pb_0$ is an effective excitation dipole moment, $\tilde\omega_0$ is the resonance frequency after accounting for the shift produced by image interaction with the film, and the $\gamma$ terms describe decay rates associated with different channels, namely, internal inelastic decay ($\gamma_{\rm in}$), decay into radiation in the homogeneous $\epsilon_1$ host medium ($\gamma_{\rm homo}$), and coupling to the film dominated by SPs ($\gamma_{\rm film}$). The latter admits the expression (see Appendix\ \ref{sec:AppB})
\begin{align}
\gamma_{\rm film}=\gamma_{\rm film}^0\,\ee^{-4\pi z_0/\lamp},
\label{gammafilm}
\end{align}
with
\begin{align}
\gamma_{\rm film}^0=\frac{\aRp}{\epsilon_1}\frac{(2\pi)^4}{\hbar\lamp^3}\ccpar{|\pb_{0\parallel}|^2/2+|p_{0z}|^2},
\nonumber
\end{align}
which combined with Eq.\ \eqref{alphafull}, and neglecting $\gamma_{\rm homo}$ as a radiative correction, allows us to write the maximum polarizability (achieved at the resonance frequency $\omega=\tilde\omega_0$) as
\begin{align}
{\rm max}\clpar{|\alpha_z|}&=\frac{1}{2}{\rm max}\clpar{|\alpha_x|}\nonumber\\&=\frac{\epsilon_1\lamp^3}{8\pi^4\aRp}\,\frac{1}{\left(1+\gamma_{\rm in}/\gamma_{\rm film}\right)}\,\ee^{4\pi z_0/\lamp}.
\label{OTfinalz0}
\end{align}
This result constitutes a generalization of Eq.\ \eqref{OTfinal}, with which it coincides for non-lossy particles ($\gamma_{\rm in}=0$) and $z_0=0$. Incidentally, in spite of the fact that our methods are based on classical electrodynamics, a factor $1/\hbar$ shows up in $\gamma_{\rm film}$, which arises when following the semiclassical prescription of dividing the dipole-emitted power by the photon energy $\hbar\omega$ in order to convert it into a rate.

The above analysis allows us to address the optimum choice of $z_0$ needed to maximize the in-coupling cross-section, which has an explicit distance dependence given by $\sigma^{\rm in-coup}_{\rm point}\propto\xi|\alpha|^2\propto\xi\,(1+\xi\,\gamma_{\rm in}/\gamma^0_{\rm film})^{-2}$, where $\xi=\ee^{4\pi z_0/\lamp}$. This expression features a maximum as a function of $\xi$ at $\xi=\gamma^0_{\rm film}/\gamma_{\rm in}$, or equivalently, $z_0=[\lamp/(4\pi)]\,\log(\gamma^0_{\rm film}/\gamma_{\rm in})$, under the condition $\gamma^0_{\rm film}>\gamma_{\rm in}$, with an increase in $\sigma^{\rm in-coup}_{\rm point}$ by a factor $\sim\gamma^0_{\rm film}/(4\gamma_{\rm in})$ relative to the touching configuration ($z_0=0$).

As an example of a realistic scenario, we note that gold and silver disk-like nanoparticles of reduced size ($<100\,$nm diameter) can already produce $\gamma^0_{\rm film}\gg\gamma_{\rm in}$ for a plasmon wavelength $\lamp=200\,$nm at visible frequencies (see Appendix\ H and \ref{InfluenceofSize}); these particles are therefore small compared with $\lamp$ ({\it i.e.}, they act as dipolar scatterers), but large enough as to undergo strong coupling to SPs, greatly exceeding the rate of internal inelastic decay. In this respect, silver is a better option than gold because $\hbar\gamma_{\rm in}$ is $\sim3$ times smaller \cite{JC1972} (21\,meV compared with 71\,meV). High-index dielectric particles offer another alternative, as they can sustain Mie resonances with $\gamma_{\rm in}=0$, so that coupling to SPs is limited by the radiative decay rate $\gamma_{\rm homo}$, which plays the same role as $\gamma_{\rm in}$ in the above analysis. Ignoring for simplicity the modification produced in $\gamma_{\rm homo}$ by the presence of the film, we can approximate it as $\gamma_{\rm homo}=4\mathfrak{f}^2\sqrt{\epsilon_1}\omega^3p_0^2/(3\hbar c^3)$ ({\it i.e.}, the emission rate from a $p_0$ dipole oscillating at frequency $\omega$ in a homogeneous $\epsilon_1$ dielectric material \cite{paper053}, where $\mathfrak{f}$ is a local-field correction factor discussed in Appendix\ \ref{sec:AppI}), from which we find an increase in in-coupling cross-section for in-plane polarization by a factor $\gamma^0_{\rm film}/(4\gamma_{\rm homo})=[3\pi\aRp/(16\mathfrak{f}^2\epsilon_1^{3/2})]\,(\lambda_0/\lamp)^3$. Multiplying this result by Eq.\ \eqref{maxsigmain}, and assuming $\mathfrak{f}\approx1$ and $A_0\approx1$ for clarity, we obtain ${\rm max}\clpar{\sigma^{\rm in-coup}_{\rm particle}}\approx3\lambda_0^2/(8\pi\epsilon_1)$, which remarkably coincides with the maximum absorption cross-section of a dipolar scatterer inside a medium of permittivity $\epsilon_1$ (incidentally this is in turn a factor of 4 smaller than the maximum possible extinction cross-section of a lossless dipolar scatterer). We conclude that, in combination with tightly focused light beams, suitably chosen noble-metal and dielectric nanoparticles can be used to increase the photon-polariton in-coupling efficiency to unity order. Additionally, resonant molecules placed in a transparent host and held at cryogenic temperatures have been shown to perform as reasonably strong two-level systems \cite{RWL12}, thus suggesting a possible realization of lossless scatterers for optimum light-SP in-coupling efficiency.
}

\begin{table*} \begin{center}
\begin{tabular}[c]{|cc|c|c|c|} \hline
&& {\color{blue} {\bf Magnitude} $=$ $(C/\aRp)\times S$} & {\color{blue} {\bf Wavelength Scaling} $S$} & {\color{blue} {\bf Prefactor} $C$} \\ \hline &&&& \\
\parbox[c]{3mm}{\multirow{4}{*}{\rotatebox[origin=c]{90}{\bf \color{blue} Point}}}
\parbox[c]{3mm}{\multirow{4}{*}{\rotatebox[origin=c]{90}{\bf \color{blue} Scatterer}}} % -------
&& $\mathrm{max}\{|\alpha_z|\}$ & $\lamp^3$ & $\epsilon_1/(8\pi^4)$ \\
&& $\mathrm{max}\{\sigma^{\rm in-coup}_{\rm point}\}$ & $\lamp^2 (\lamp/\lambda_0)$ & $2 \sqrt{\epsilon_1} A_0/\pi^2$ \\
&& $\mathrm{max}\{\sigma^{\rm out-coup}_{\rm point}\}$  & $\lamp (\lamp/\lambda_0)^3$ & $4 \epsilon_1^{3/2}/(3\pi^2)$  \\
&& $N_{\rm polariton}/N_{\rm photon}$ & $(\lamp/\lambda_0)^3$ & $4 g \epsilon_1^{3/2}/\pi$ \\ &&&& \\ \hline  &&&& \\
\parbox[c]{3mm}{\multirow{3}{*}{\rotatebox[origin=c]{90}{\bf \color{blue} Line}}}
\parbox[c]{3mm}{\multirow{3}{*}{\rotatebox[origin=c]{90}{\bf \color{blue} Scatterer}}} % -------
&& $\mathrm{max}\{|\mathcal{A}_z|\}$ & $\lamp^2$ & $\epsilon_1/(8\pi^3)$\\
&& $\mathrm{max}\{\sigma^{\rm in-coup}_{\rm line}\}$ & $\lamp (\lamp/\lambda_0)$ & $\sqrt{\epsilon_1}/\pi$ \\
&& $N_{\rm polariton}/N_{\rm photon}$ & $(\lamp/\lambda_0)^2$ & $g \epsilon_1/\pi$ \\ &&&& \\ \hline
\end{tabular} \end{center} \label{Table2} \caption{Compilation of wavelength scalings for several relevant coupling magnitudes.} \end{table*}

\begin{figure*}[htbp]
\begin{centering}
\includegraphics[width=0.9\textwidth]{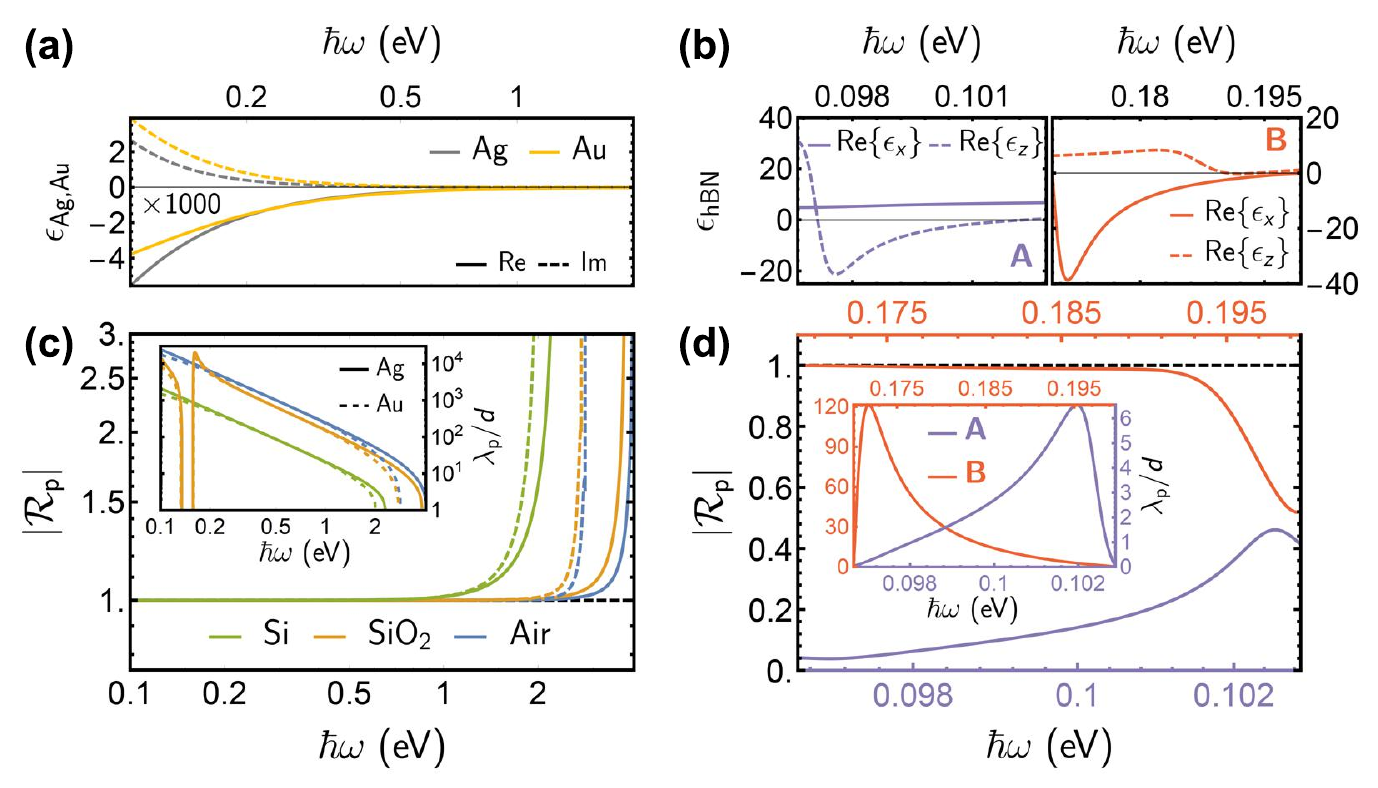}
\par\end{centering}
\caption{Film parameters for noble metals and hBN. We show the film-material dielectric functions (a,b), the correction factor $\aRp$ (c,d), and the $\lamp/d$ ratio (insets to (c,d)) for gold and silver (a,c), and hBN (b,d). The latter is plotted for self-standing films over the two narrow mid-infrared spectral ranges exhibiting hyperbolic behavior ($\Ree\clpar{\epsilon_x}\Ree\clpar{\epsilon_z}<0$). Noble metal films in (a,c) are considered to be fully embedded in silicon, silica, or air (see legend in (c)). We take the dielectric functions of hBN \cite{GPR1966}, silica \cite{P1985}, and silicon \cite{L1980} from tabulated experimental data, {\color{bluecol} while we use the Drude-like Eq.\ \eqref{Drudemetal} for gold and silver}.}
\label{Fig4}
\end{figure*}

\section{CONCLUDING REMARKS}

The analytical limits to the polarizability of point and line scatterers formulated above (see summary in Table\ \ref{Table2}) impose a severe restriction on the ability of small elements to couple light and polaritons. These quantities have a smooth dependence on the dielectric environment, including a correction factor $\aRp$ that we plot in Fig.\ \ref{Fig4} for plasmons in thin silver films and hyperbolic phonon polaritons in hBN.

The combination of these scatterers with other elements can boost coupling by enhancing the field intensity to which they are exposed. This is the case of metallic tips, which are extensively used in scanning near-field optical microscopy to yield spectrally-filtered images of polaritons and their interactions with structured films \cite{BKO10,FAB11,MH14,WAL16}. Indeed, roughly modeling the tip as a perfect-conductor ellipsoid, the intensity enhancement at the tip apex scales as $(1+{\rm AR})^2$ with its aspect ratio (AR =length/diameter, typically $\gg1$), a factor that must be multiplied by $\sigma^{\rm in-coup}_{\rm point}$ to produce the actual in-coupling cross-section in this configuration. Our results also impose severe restrictions on in-plane polariton optics. Remarkably, the maximum extinction and scattering cross-sections of small scatterers are $\sim\lamp$, independent of film thickness and dielectric environment.

As a viable alternative, our calculations support the use of {\color{bluecol} resonant nanoparticles placed at an optimum distance from the film in order to boost the in-coupling efficiency, which for an optimally designed scatterer could reach a coupling conversion fraction $N_{\rm photon}/N_{\rm polariton}\sim1$ for nearly lossless particles}. This specific proposal should be readily attainable with currently available nanofabrication technology, either using lithographycally patterned high-index resonators or colloids deposited on the targeted SP-supporting film.

{\color{bluecol} Alternative methods to excite SPs in thin films appear at a disadvantage compared with this proposal. In particular, one could use photoluminescent particles ({\it e.g.}, quantum dots), which couple preferentially to SPs rather than radiation when placed near a SP-supporting film ({\it i.e.}, $\gamma_{\rm film}\gg\gamma_{\rm homo}$); however, the emitted SPs are incoherent relative to the light used for primary excitation, in contrast to the SP field produced by the elastic scattering here investigated; additionally, the excitation cross-section needed to produce photoluminescence can be strongly reduced by broadening of the targeted initial particle transition due to interaction with the film, thus resulting in a much poorer photon-to-SP coupling efficiency.}

In brief, our results provide some basic tools to understand and quantify light-polariton coupling and in-plane polariton scattering in 2D materials, while they allow us to explore optimum coupling geometries, such as the resonant particle proposed above. An extension to include retardation becomes necessary when the polariton wavelength is not much smaller than the light wavelength, for which we essentially follow the same analytical procedure as above (see Appendix\ \ref{sec:Retardation} for more details). However, the degree of confinement is reduced in this regime, characterized by loosely bound SP modes that are less amenable to the applications discussed in the introduction. Multipolar scatterers constitute another useful extension, for which we expect higher bounds to the light-polariton coupling, increasing with the multipolar order \cite{MT18}; indeed, these elements should couple less efficienty {\color{bluecol} (and therefore decay more slowly) to polaritons, and consequently}, their SP-limited width should be smaller, translating into higher bounds to the cross sections under investigation. The combination of several point or line scatterers provides a practical recipe to produce multipolar scatterers, and we anticipate that such arrays could indeed reach larger light-polariton coupling by exploiting lattice resonances at the in-plane polariton wavelength \cite{paper090}.

\section*{APPENDIX}

\appendix

\section{ Scattered SP Field Produced by a Point Dipole near a SP-Supporting Film} 
\label{sec:AppA}

The electric field produced by a dipole $\pb$ placed at $\rb'$ inside a homogeneous medium of permittivity $\epsilon_1$ reduces in the quasistatic limit to $\Eb^{\rm homo}=-\nabla\phi^{\rm homo}$, where \cite{J99}
\begin{align}
\phi^{\rm homo}&=\frac{1}{\epsilon_1} (\pb\cdot\nabla') \frac{1}{|\rb-\rb'|}\nonumber\\&=\frac{1}{\epsilon_1} \pb\cdot\nabla'\int \frac{d^2\kparb}{2\pi\kpar} \ee^{\ii\kparb\cdot(\Rb-\Rb')-\kpar|z-z'|}, \label{Edir}
\end{align}
is the electrostatic potential, we use the notation $\rb=(\Rb,z)$ with $\Rb=(x,y)$, and the rightmost expression stems from the 2D momentum representation of the Coulomb potential. In the presence of a planar interface at $z=0$, we find it useful to relate the reflected field to the Fresnel coefficient for p polarization $\rp(\kpar,\omega)$. For this purpose, we consider a p-polarized electromagnetic plane wave \cite{paper052} $(\kparb k_{z1}+\kpar^2\zz)\,\ee^{\ii\kparb\cdot\Rb-\ii k_{z1}z}$ incident from $z>0$ and its reflection $\rp\,(-\kparb k_{z1}+\kpar^2\zz)\,\ee^{\ii\kparb\cdot\Rb+\ii k_{z1}z}$. In the quasistatic limit ($c\rightarrow\infty$), we have $k_{z1}=\sqrt{\epsilon_1\omega^2/c-\kpar^2}\rightarrow\ii\kpar$, which permits us to write the sum of incident and reflected fields as $-\nabla\phi$ in terms of a potential $\phi=-\kpar\,(\ee^{\kpar z}-\rp\ee^{-\kpar z})\,\ee^{\ii\kparb\cdot\Rb}$. We conclude that each component $\ee^{\ii\kparb\cdot(\Rb-\Rb')+\kpar (z-z')}$ in the {\it incident} potential of the dipole (Eq.\ \eqref{Edir}) generates a {\it reflected} potential $-\rp\,\ee^{\ii\kparb\cdot(\Rb-\Rb')-\kpar (z+z')}$. Performing this substitution in Eq.\ \eqref{Edir}, we find the reflected potential in the $z>0$ region
\begin{align}
\phi^{\rm ref}&=\frac{-1}{\epsilon_1} (\pb\cdot\nabla') \int \frac{d^2\kparb}{2\pi\kpar}\,\ee^{\ii\kparb\cdot(\Rb-\Rb')-\kpar (z+z')}\,\rp \nonumber\\
&=\frac{1}{\epsilon_1} (\pb_\parallel\cdot\nabla_\Rb-p_z\,\partial_z) \int \frac{d^2\kparb}{2\pi\kpar}\,\ee^{\ii\kparb\cdot\Rb-\kpar (z+z_0)}\,\rp, \label{Eref1}
\end{align}
{\color{bluecol} where the second line is obtained by first making the replacement $\nabla'\rightarrow-\nabla_\Rb+\zz\,\partial_z$  and then specifying the dipole position at $\rb'=(0,0,z_0)$.} We now argue that SPs are signaled by a divergence of $\rp$ at $\kpar=\kparsp$, so we use Eq.\ \eqref{rp} to work out the $\kparb$ integral analytically, leaving us with a dimensionless factor $\Rp$ that is a smooth function of $\kpar$ and can be approximated by its value at $\kpar=\kparsp$ (see Eq.\ \eqref{rp}). Indeed, any $\kpar$-independent term in $\rp$ contributes as $\propto1/|\rb-\rb'|$ to the integral in Eq.\ \eqref{Eref1}, or equivalently, an electric field that decays as $\sim1/R^3$ at large in-plane distances $R$; as we are interested in surface modes with slower decay $\sim1/\sqrt{R}$, such terms can be dismissed, and we further argue that any divergence-free term in $\rp$ will contribute negligibly at large distances. Integration over the angle of $\kparb$ in Eq.\ \eqref{Eref1} produces a Bessel function $J_0(\kpar R)$ that in the $|\kparsp R|\gg1$ limit leads to the asymptotic approximation \cite{AS1972}
\begin{align*}
&\int \frac{d^2\kparb}{2\pi\kpar}\,\ee^{\ii\kparb\cdot\Rb-\kpar(z+z_0)}\,\rp \\
&\!\approx\Rp\kparsp\!\!\int_0^\infty \!\!\!\!\frac{d\kpar}{\kpar-\kparsp}\frac{\ee^{-\kpar(z+z_0)}}{\sqrt{2\pi\kpar R}}\!\!\left[\ee^{\ii(\kpar R-\pi/4)}\!+\!\ee^{-\ii(\kpar R-\pi/4)}\!\right]\!,
\end{align*}
to which only non-resonant terms $\propto1/r^3$ are incorporated by extending the range of integration down to $-\infty$ and replacing $\sqrt{\kpar^2+0^+}$ for $\kpar$ in $\ee^{-\kpar z}$; then, by closing the contour in the upper (for the $\ee^{\ii\kpar R}$ term) or lower (for $\ee^{-\ii\kpar R}$) complex plane in $\kpar$, and neglecting non-resonant contributions from branching points, the $\rp$ pole (in the upper plane because $\Imm\clpar{\kparsp}>0$) is found to contribute only through the outward wave as
\begin{align*}
&\int \frac{d^2\kparb}{2\pi\kpar}\,\ee^{\ii\kparb\cdot\Rb-\kpar(z+z_0)}\,\rp\\
&\approx\ii\Rp\kparsp\sqrt{\frac{2\pi}{\kparsp R}}\,\ee^{\kparsp(\ii R-z-z_0)-\ii\pi/4}.
\end{align*}
Finally, using this expression in Eq.\ \eqref{Eref1}, performing the spatial derivatives {\it via} the substitution $\nabla=\ii\kparsp\RR$ ({\it i.e.}, neglecting terms that decay faster than $1/\sqrt{R}$), taking $z_0=0$, and multiplying by $-\nabla=-\ii\kparsp(\RR+\ii\zz)$ to compute the scattered electric field from the potential, we readily obtain Eq.\ \eqref{Eprad}. We note that for finite dipole-film separation $z_0$ the field carries an additional factor $\ee^{-\kparsp z_0}$.

\section{Scattered SP Power Produced by a Point Dipole} 
\label{sec:AppB}

Neglecting inelastic losses in the film, the power carried by the scattered SP field must be equal to $P^{\rm scat}=\hbar\omega\Gamma_{\rm film}$, where $\Gamma_{\rm film}$ is the dipole decay rate. In a homogeneous dielectric, the decay rate ($\Gamma_{\rm homo}\propto1/c^3$) is dominated by coupling to radiation and vanishes in the quasistatic limit under consideration, but in the vicinity of a film the reflected field provides a nonzero contribution arising from coupling to SPs. More precisely, we have \cite{paper053} $\Gamma_{\rm film}=(2/\hbar){\rm Im}\{\pb^*\cdot\Eb^{\rm ref}\}$, where $\Eb^{\rm ref}=-\nabla\phi^{\rm ref}$ is evaluated at the position of the dipole $\rb=\rb'=(0,0,z_0)$. Using Eq.\ \eqref{Eref1} for $\phi^{\rm ref}$ with $\rb=\rb'=0$, we find \cite{FW1984}
\begin{align}
\Gamma_{\rm film}=\frac{2}{\epsilon_1\hbar}\left(|\pb_\parallel|^2/2+|p_z|^2\right)\int_0^\infty \kpar^2\,d\kpar\,\ee^{-2\kpar z_0}\,\Imm\clpar{\rp}.
\label{Gamma}
\end{align}
Approximating $\rp$ by the pole contribution of the dominant SP (plasmon-pole approximation, see Eq.\ \eqref{rp}), and noting that in the limit of negligible inelastic losses $\kparsp$ must have an infinitesimal positive imaginary part, we find $\Imm\clpar{\rp}\approx\pi\aRp\kparsp\delta(\kpar-\kparsp)$, which upon insertion into Eq.\ \eqref{Gamma} and multiplication by $\hbar\omega$ directly yields
\begin{align}
P^{\rm scat}=\frac{2\pi\aRp}{\epsilon_1}\,\ee^{-2\kparsp z_0}\,\omega\kparsp^3\,\left(|\pb_\parallel|^2/2+|p_z|^2\right),
\label{PscatA}
\end{align}
and from here, with $\kparsp\approx2\pi/\lamp$ and $z_0=0$, we obtain Eq.\ \eqref{Pscat}.

\section{Power Flux Associated with a SP Plane Wave} 
\label{sec:AppC}

We intend to find the power flux $I_{\rm p}$ carried by a SP plane wave (see Eq.\ \eqref{Ep}), defined as the propagated power per unit of transversal length. Obviously, this quantity must be proportional to $|E_0|^2$; we can derive a general expression for it that applies to any type of film by considering the field scattered by a dipole (Eq.\ \eqref{Eprad}), approximating the circular SP wavefront as a plane wave for each direction $\Rb$ in the $|\kparsp R|\gg1$ limit, and integrating over in-plane scattering directions. More precisely, we have
\begin{align}
P^{\rm scat}=I_{\rm p}\int R\,d\varphi \frac{1}{|\kparsp R|} \left|E^{\rm scat}_{\RR}/E_0\right|^2,
\nonumber
\end{align}
where $R\,d\varphi$ is the element of transversal length in the circular wavefront and we replace the PW intensity $|E_0|^2$ by the scattered field intensity $|\Eb^{\rm scat}|^2$ ({\it cf.} Eqs.\ \eqref{Ep} and \eqref{Eprad}). Inserting Eq.\ \eqref{Eprad} in this expression, we find
\begin{align}
P^{\rm scat}=\left(\frac{2\pi\aRp}{\epsilon_1}\right)^2
\frac{I_{\rm p}}{|E_0|^2}\,\kparsp^5\,\ccpar{|\pb_\parallel|^2/2+|p_z|^2},
\nonumber
\end{align}
and comparison with Eq.\ \eqref{PscatA} finally yields $I_{\rm p}=\epsilon_1\omega|E_0|^2/(2\pi\aRp\kparsp^2)$, and from here Eq.\ \eqref{Ip} with $\kparsp\approx2\pi/\lamp$.

\section{Surface-Polariton Extinction Cross-Section} 
\label{sec:AppD}

We obtain the extinction cross-section for an incident SP plane wave (Eq.\ \eqref{Ep}) by integrating the total SP field intensity for large $|\kparsp R|$ near the forward scattering direction. In fact, we only need to consider the out-of-plane field component right outside the film, as the in-plane component has the same intensity and is just $-\pi/2$ out of phase, while the $z$ dependence is common for incident and scattered waves. The integrated field intensity is then given by
\begin{align}
\int dy \left|E_0\ee^{\ii\kparsp x}+E^{\rm scat}_{\RR}\frac{\ee^{\ii\kparsp R}}{\sqrt{\kparsp R}}\right|^2.
\nonumber
\end{align}
The square of the $E_0$ term in this expression is independent of $y$ (the in-plane coordinate transversal to the propagation direction $x$), and therefore, it produces a contribution $L|E_0|^2$ proportional to the lateral beam size $L$. The square of the second term was considered above, integrated over scattering directions $\RR$ to yield the scattered power; however, near the forward direction, this term has a $1/R$ dependence that makes it vanish in the $|\kparsp R|\gg1$ limit. The remaining interference term can be understood as the portion of incident beam that is extincted by interaction with the scatterer; when normalized to $|E_0|^2$, it should give a negative length that is subtracted from the $L$ contribution of the direct beam, and consequently, it represents an extinction cross-section with units of length, which we thus write as
\begin{align}
\sigma^{\rm ext}_{\rm point}=-2\,\Ree\clpar{\int dy \frac{E^{\rm scat}_{\RR}}{E_0}\frac{\ee^{\ii\kparsp (R-x)}}{\sqrt{\kparsp R}}}.
\nonumber
\end{align}
Noticing that the exponential term in this integral produces fast sign cancellations in the far-field limit ($|\kparsp R|\gg1$) unless $R\approx x$, we can approximate $R-x\approx y^2/2x$ in the exponent, as well as $R\approx x$ and $\RR\approx\xx$ in the rest of the expression. Additionally, we set the dipole induced by the incident SP plane wave (see Eq.\ \eqref{Ep}) as $\pb=E_0(\alpha_x\xx+\ii\alpha_z\zz)$, which permits writing (see Eq.\ \eqref{Eprad})
\begin{align}
&\sigma^{\rm ext}_{\rm point}=\nonumber\\
&\frac{2}{\epsilon_1}\,\Imm\clpar{\ee^{-\ii\pi/4}\,\Rp\,\kparsp^3\; (\alpha_x+\alpha_z)\int dy\,\sqrt{\frac{2\pi}{\kparsp x}}\,\ee^{\ii\kparsp y^2/2x}}.
\nonumber
\end{align}
With the change of variable $\theta=y\sqrt{\kparsp/(2x)}$, we are left with the integral \cite{GR1980} $\int \!d\theta\,\exp(\ii\theta^2)=\sqrt{\pi}\,\ee^{\ii\pi/4}$, so the above expression directly leads to Eq.\ \eqref{sigmaext}, where we approximate $\Rp\approx\aRp$ and $\kparsp\approx2\pi/\lamp$ under the assumption of negligible film-material losses. Incidentally, for finite scattered-film separation $z_0$, the above expression is simply corrected by a factor $\ee^{-\kparsp z_0}$.

\section{Scattered SP and Light Fields Produced by a Line Scatterer} 
\label{sec:AppE}

Each dipole element in a line scatterer extending along $y$ contributes to the SP field $-\nabla\phi^{\rm ref}$ with a potential $\phi^{\rm ref}$ given by Eq.\ \eqref{Eref1}, in which $\vec{\mathcal{P}}dy'$ (the dipole element along the line) must be substituted for $\pb$. The total field produced by the entire line of dipole elements is then obtained by integrating over $y'$, which produces $\delta(k_y)$ and directly leads to
\begin{align}
\Eb^{\rm scat}=\frac{1}{\epsilon_1} (\pb\cdot\nabla')\nabla \int \frac{dk_x}{|k_x|}\,\ee^{\ii k_x(x-x')-|k_x|(z+z')}\,\rp. \label{Escat2}
\end{align}
In the spirit of the plasmon-pole approximation (Eq.\ \eqref{rp}), we write $\rp\approx2\Rp |k_x|\kparsp/(k_x^2-\kparsp^2)$, which is invariant under sign changes of $k_x$ and has the same pole structure as Eq.\ \eqref{rp}. Using this in Eq.\ \eqref{Escat2} and integrating in the complex plane for $k_x$, we find the scattered field to reduce to a plane wave like Eq.\ \eqref{Ep} with $E_0$ replaced by $E^{\rm scat}_{\rm line}=[8\pi^3\Rp/(\epsilon_1\lamp^2)]\,(\ii\mathcal{P}_x+\mathcal{P}_z)$.

A similar analysis can be carried out for the upward light emission from the line of dipoles, each of them contributing with a field proportional to $k^2[\vec{\mathcal{P}}-(\nabla\cdot\vec{\mathcal{P}})\nabla]\ee^{\ii k_1 r}/r$, where $k=\omega/c$ and $k_1=k\sqrt{\epsilon_1}$. Integrating over $y$, this expression yields a field $k^2[\vec{\mathcal{P}}-(\nabla\cdot\vec{\mathcal{P}})\nabla]\ee^{\ii k_1 R+\ii\pi/4}\sqrt{2\pi/(k_1R)}$, where $R=\sqrt{x^2+z^2}$. Finally, integrating the Poynting vector resulting from this field over the upper hemisphere of emission directions, we finally get a radiated power $(2\pi^3\omega/\lambda_0^2)L\,|{\bf\mathcal{P}}|^2$, which we use in the main text to calculate the polariton-to-photon coupling efficiency by a line scatterer.

\section{Polariton Dispersion Relation and Reflection Coefficients}
\label{sec:AppF}

A thin film of finite thickness with low-loss metallic permittivity ({\it i.e.}, ${\rm Re}\{\epsilonm\}<0$ and ${\rm Im}\{\epsilonm\}\ll-{\rm Re}\{\epsilonm\}$) supports plasmons, the dispersion relation of which is determined by the Fabry-Perot condition $1 - r_{{\rm p,m}1}r_{{\rm p,m}2}\,\ee^{2 \ii k_{z{\rm m}} d}=0$, where $r_{{\rm p,m}j}=(\epsilon_j k_{z{\rm m}}-\epsilonm k_{zj})/(\epsilon_j k_{z{\rm m}}+\epsilonm k_{zj})$ is the Fresnel coefficient for reflection of p waves at the planar interface formed by the metal and the surrounding medium $j$, while $k_{zj}=\sqrt{k^2\epsilon_j-\kpar^2}$ is the normal light wave vector in medium $j$. We are interested in the $\lamp\ll\lambda_0$ limit, so we neglect retardation, which allows us to write $r_{{\rm p,m}j}\approx(\epsilon_j-\epsilonm)/(\epsilon_j+\epsilonm)$ and $k_{zj}\approx\ii\kpar$, and from here the plasmon dispersion relation reduces to
\begin{align}
\kparsp = \frac{1}{2d} \log \sqpar{\frac{(\epsilon_1-\epsilonm)(\epsilon_2-\epsilonm)}{(\epsilon_1+\epsilonm)(\epsilon_2+\epsilonm)}},
\nonumber
\end{align}
which is recast in terms of the polariton wavelength in Eq.\ \eqref{lampd} and reduces to Eq.\ \eqref{kparw} in the zero-thickness limit ($d\rightarrow0$, $|\epsilonm|\rightarrow\infty$, $\epsilonm d\rightarrow4\pi\ii\sigma/\omega$). The reflection residue $\Rp$ is obtained from the Fresnel reflection coefficient of the film from medium 1
\begin{align}
\rp=r_{{\rm p},1{\rm m}}+\frac{t_{{\rm p,1}m}t_{{\rm p,m}1}r_{{\rm p,m}2}\ee^{2 \ii k_{z{\rm m}} d}}{1 - r_{{\rm p,m}1}r_{{\rm p,m}2} \ee^{2 \ii k_{z{\rm m}} d}},
\label{rpF}
\end{align}
where $t_{{\rm p,m}j}=2\sqrt{\epsilon_j\epsilonm}k_{z{\rm m}}/(\epsilon_j k_{z{\rm m}}+\epsilonm k_{zj})\approx2\sqrt{\epsilon_j\epsilonm}/(\epsilon_j+\epsilonm)$ and $t_{{\rm p},j{\rm m}}=(k_{zj}/k_{z{\rm m}})t_{{\rm p,m}j}\approx2\sqrt{\epsilon_j\epsilonm}/(\epsilon_j+\epsilonm)$ are the in and out transmission coefficients at the $j$m interface. We then substitute $r_{{\rm p,m}1}r_{{\rm p,m}2}$ by $\ee^{-2 \ii k_{z{\rm m}} d}$ evaluated at $\kpar=\kparsp$ ({\it i.e.}, the noted Fabry-Perot condition for the SP mode), and expand everything to first order in $\kpar-\kparsp$ to produce the result shown in Eq.\ \eqref{Rphomo} and Table\ \ref{Table1}.

For completeness, we note that the film reflection coefficient for s polarization is given by Eq.\ \eqref{rpF} with p replaced by s, $r_{{\rm s,m}j}=(k_{z{\rm m}}-k_{zj})/(k_{z{\rm m}}+k_{zj})$,
$t_{{\rm s,m}j}=2k_{z{\rm m}}/(k_{z{\rm m}}+k_{zj})$, and $t_{{\rm s},j{\rm m}}=2k_{zj}/(k_{z{\rm m}}+k_{zj})$, and in the quasistatic limit we have $r_{{\rm s,m}j}\approx0$ and $t_{{\rm s,m}j},\,t_{{\rm s},j{\rm m}}\approx1$.

For anisotropic films characterized by permittivities $\epsilon_x=\epsilon_y$ and $\epsilon_z$ for in- and out-of-plane polarization, respectively, the above results need to be amended by (1) replacing $\epsilonm$ by $\epsilon_x$ and (2) taking $k_{z{\rm m}}=\sqrt{(k^2-\kpar^2/\epsilon_z)\epsilon_x}$ for p polarization and $k_{z{\rm m}}=\sqrt{k^2\epsilon_x-\kpar^2}$ for s polarization. An analysis similar to the homogeneous metallic film leads to the expressions for $\lamp$ and $\Rp$ given in Table\ \ref{Table1}.

\section{Maximum Field Produced by a Focused Light Beam}
\label{sec:AppG}

In order to quantify the maximum photon-to-polariton conversion efficiency, we first need to determine the ratio $|E_x|^2/P^{\rm beam}$ between the light intensity acting on the scatterer and the power $P^{\rm beam}$ carried by the focused beam. We take the beam to be in a homogeneous medium of permittivity $\epsilon_1$ and consider a point scatterer at the focal point with in-plane polarization along $x$. The beam can be constructed as the superposition of plane waves of polarization $\sigma$ ($=$s, p), parallel wave vector $\kparb$, and unit polarization vector $\eh^{1+}_{\kparb\sigma}$, with amplitudes $a_{\kparb\sigma}$ such that the focal electric field reduces to $\Eb=\sum_\sigma\int_{\kpar<k_1} d^2\kparb a_{\kparb\sigma} \eh^{1+}_{\kparb\sigma}$, where $k_1=k\sqrt{\epsilon_1}$ and $k=\omega/c$. Also, the beam power is readily obtained by integrating the associated Poynting vector over a plane perpendicular to the beam, which yields $P^{\rm beam}=[c\sqrt{\epsilon_1}/(2\pi)]\sum_\sigma\int_{\kpar<k_1} d^2\kparb \sqabs{a_{\kparb\sigma}}$. A Gaussian beam corresponds to the choice of coefficients \cite{NH06} $a_{\kparb\sigma}=\xx\cdot\eh^{1+}_{\kparb\sigma}\sqrt{k_{z1}/k_1}\,\ee^{-\beta\kpar^2/k^2}$, where $\beta$ controls the degree of focusing. The field is maximized for $\beta=0$, which produces a ratio $|E_x|^2/P^{\rm beam}\approx0.4646\,k^2\sqrt{\epsilon_1}/c$. However, this is not the optimum solution that yields the maximum possible value of the ratio under consideration. Direct application of the Lagrange multiplier method yields the optimum choice $a_{\kparb{\rm s}}\propto-k_yk_1/(\kpar k_{z1})$ and $a_{\kparb{\rm p}}\propto k_x/\kpar$, which upon insertion in the above integrals yields $|E_x|^2/P^{\rm beam}=(2/3)k^2\sqrt{\epsilon_1}/c$.

Incidentally, a similar procedure can be followed to include the effect of reflection by the film, simply by multiplying the coefficients $a_{\kparb\sigma}$ in the expression for $E_x$ by either $1+\rs$ or $1-\rp$ for $\sigma=\,$s or p polarization, respectively, which involves materials and thickness-dependent reflection coefficients $r_\sigma$. For simplicity, we ignore this effect in our calculations of the $|E_x|^2/P^{\rm beam}$ ratio.

We now repeat the above procedure for a line scatterer with translational invariance along $y$ by considering only p waves with parallel wave vectors $\kparb\parallel\xx$. The ratio of the focal-spot intensity to the beam power now becomes $|E_x|^2L/P^{\rm beam}=0.1887\,k/c$ and $|E_x|^2L/P^{\rm beam}=k/4c$ for Gaussian and optimally-focused beams, where $L$ is the normalization scatterer length along $y$.

\section{Effective Polarizability of a Particle near a Film} 
\label{sec:AppH}

The $3\times3$ polarizability tensor of a particle characterized by a resonance frequency $\omega_0$ can be approximated as \cite{PN1966}
{\color{bluecol}
\begin{align}
\alpha_{\rm part}=\frac{\hbar^{-1}\,\pb_0\otimes\pb_0}{\omega_0-\omega-\ii(\gamma_{\rm in}+\gamma_{\rm homo})/2}
\label{alphapart}
\end{align}
when placed in a homogeneous environment in the absence of the film, where $\pb_0$ acts as an {\color{bluecol} {\it effective}} resonance transition dipole and we have split the decay rate into two components, namely, $\gamma_{\rm in}$ for intrinsic inelastic processes and $\gamma_{\rm homo}=4\mathfrak{f}^2\sqrt{\epsilon_1}\omega^3p_0^2/(3\hbar c^3)$ for coupling to radiation in the surrounding $\epsilon_1$ material \cite{paper053} (see discussion on the local-field correction factor $\mathfrak{f}$ below). The dipole induced on the particle in response to an external electric field $\Eb^{\rm ext}$ is simply given by $\pb=\alpha_{\rm part}\Eb^{\rm ext}$. This expression holds even when the particle is near a film surface, provided we substitute $\Eb^{\rm ext}$ by the sum of incident and surface-reflected fields. The latter consists of the reflection of both the incident field and the field that is self-consistently generated by the dipole. We seek to obtain an effective polarizability for the particle in the presence of the film, defined in accordance with the convention used in the main text, in such as way that it gives the induced dipole when multiplied by the sum of incident and specularly reflected fields ({\it i.e.}, the reflection of the field produced by the induced dipole is already included in the polarizability). We denote this field sum as $\Eb^{\rm inc}$ (incidentally, the coefficients defined in Eq.\ \eqref{AA} is meant to compensate precisely for the difference between $\Eb^{\rm ext}$ and $\Eb^{\rm inc}$), and write the self-consistent dipole as $\pb=\alpha(\Eb^{\rm inc}+\mathfrak{G}\cdot\pb)$, where $\mathfrak{G}\cdot\pb$ is the additional surface-reflected component produced by the dipole and acting back on itself, which is obviously proportional to $\pb$ through a $3\times3$ matrix $\mathfrak{G}$. The effective polarizability tensor that relates $\pb$ to $\Eb^{\rm inc}$ then reduces to
\begin{align}
\alpha=\frac{1}{\alpha_{\rm part}^{-1}-\mathfrak{G}}.
\label{alphaeff}
\end{align}
We readily obtain  $\mathfrak{G}$ from the surface reflection of the dipole field $-\nabla\phi^{\rm ref}$ (see Eq.\ \eqref{Eref1}), which leads to
\begin{align}
\mathfrak{G}=G\,
\begin{bmatrix} 1/2 & 0 & 0 \\ 0 & 1/2 & 0 \\ 0 & 0 & 1 \end{bmatrix} \nonumber
\end{align}
with
\begin{align}
G=\frac{1}{\epsilon_1}\int_0^\infty \kpar^2\,d\kparb\,\rp\,\ee^{-2\kpar z_0}. \nonumber
\end{align}
Now, introducing Eq.\ \eqref{alphapart} into Eq.\ \eqref{alphaeff} and assuming $\pb_0$ to be oriented either parallel or perpendicular to the film, the effective polarizability can be rewritten as shown in Eq.\ \eqref{alphafull}, where the resonance frequency $\tilde\omega_0=\omega_0-\hbar^{-1}\Ree\{G\}\,\left(|\pb_{0\parallel}|^2/2+|p_{0z}|^2\right)$ is shifted due to the dipole image potential captured in $\Ree\{G\}$, while $\gamma_{\rm film}=2\hbar^{-1}\,\Imm\{G\}\,\left(|\pb_{0\parallel}|^2/2+|p_{0z}|^2\right)$ coincides with decay rate discussed in Eq.\ \eqref{Gamma}, here specified for a dipole strength $\pb_0$ ({\it i.e.}, the effective particle excitation dipole, rather than the field-induced dipole). Using again the plasmon-pole approximation for $\rp$ ({\it i.e.}, assuming a negligibly small imaginary part of $\kp$ and writing ${\rm Im}\{\rp\}\approx\pi\Rp\kparsp\delta(\kpar-\kparsp)$ from Eq.\ \eqref{rp}), we obtain the result given in Eq.\ \eqref{gammafilm} of the main text, which coincides with $P^{\rm scat}/(\hbar\omega)$ for $z_0=0$ (see Eq.\ \eqref{Pscat}, with $\pb_0$ substituted for $\pb$). At resonance ($\omega=\tilde\omega_0$), we readily find from Eq.\ \eqref{alphafull} a maximum purely imaginary effective polarizability
\begin{align}
\max\clpar{|\alpha|}=\frac{2}{\hbar}\,p_0^2\,(\gamma_{\rm in}+\gamma_{\rm homo}+\gamma_{\rm film})^{-1}.
\label{maxalphafilm}
\end{align}
In what follows, we neglect radiative decay under the assumption that it is small compared with intrinsic losses ({\it i.e.}, $\gamma_{\rm homo}\gg\gamma_{\rm film}$). Reassuringly, from Eqs.\ \eqref{gammafilm} and \eqref{maxalphafilm} we recover the limit of Eq.\ \eqref{OTfinal} in the absence of intrinsic losses ($\gamma_{\rm in}=0$), while for lossy particles we obtain Eq.\ \eqref{OTfinalz0} (see main text), which predicts a maximum polarizability reduced by a factor $1/(1+\gamma_{\rm in}/\gamma_{\rm film})$; if coupling to SPs in the film is large compared to intrinsic inelastic decay ({\it i.e.}, $\gamma_{\rm film}\ll\gamma_{\rm in}$), the particle behaves as a lossless scatterer. We are therefore interested in studying the fraction $\gamma_{\rm film}/\gamma_{\rm in}$ in a realistic scenario.

We focus on metal nanoparticles hosting a spectrally-isolated plasmon resonance $j$ at a light wavelength that is sufficiently large compared with the particle size as to neglect retardation effects. The particle polarizability (for clarity, we discuss a scalar component along the polarization direction) then reduces to \cite{paper300}
\begin{align}
\alpha=\frac{\epsilon_1}{4\pi}
\frac{V_j}{(\epsilon_{\rm m}/\epsilon_1-1)^{-1}-(\varepsilon_j-1)^{-1}},
\label{epsilonES}
\end{align}
where $\epsilon_{\rm m}$ is the metal permittivity, while $V_j$ and $\varepsilon_j$ are a volume and (real) permittivity eigenvalue associated with the specific mode $j$. We now consider a plasmonic metal described through a permittivity \begin{align}
\epsilon_{\rm m}=\epsilon_{\rm b}-\frac{\wp^2}{\omega(\omega+\ii\gamma_{\rm in})},
\label{Drudemetal}
\end{align}
where $\epsilon_{\rm b}$ (taken as 1 in Fig.\ \ref{Fig1} for simplicity) is introduced to account for inner-shell electron polarization ({\it e.g.}, $\epsilon_{\rm b}=9.5$ and 4.0 in gold and silver, respectively). The other two parameters can be approximated as \cite{JC1972} $\hbar\wp=9.06\,$eV and $\hbar\gamma_{\rm in}=71\,$meV ($\hbar\wp=9.17\,$eV and $\hbar\gamma_{\rm in}=21\,$meV) in gold (silver). Upon insertion of this permittivity into Eq.\ \eqref{epsilonES}, we find a polarizability given by Eq.\ \eqref{alphapart} with a resonance frequency $\omega_0=\wp/\sqrt{\epsilon_{\rm b}-\epsilon_1\varepsilon_j}$ and an effective excitation dipole
\begin{align}
p_0=\frac{\epsilon_1(1-\varepsilon_j)}{\epsilon_{\rm b}-\epsilon_1\varepsilon_j}\sqrt{\frac{\hbar\omega_0V_j}{8\pi}}.
\label{p0}
\end{align}
In the derivation of this result, we have neglected a frequency-independent term that is usually small compared with the plasmon contribution. The sum of $V_j$ for all modes $j$ is known to be equal to the particle volume $V$ \cite{paper300}, but we consider for simplicity an oblate ellipsoid with rotational axis along $z$ and polarization along its diameter; then, the particle only hosts one dipolar mode $j$, the parameters of which are \cite{J1945} $V_j=V$ and $\varepsilon_j\equiv\varepsilon=1-1/L<-1$, where $L=(r^2/2)\Delta^-3\left[\pi/2-\arctan(1/\Delta)-\Delta/r^2\right]$ is the depolarization factor, $r>1$ is the diameter-to-height aspect ratio, and $\Delta=\sqrt{r^2-1}$. We are now prepared to evaluate $\gamma_{\rm film}/\gamma_{\rm in}$ with the help of Eq.\ \eqref{p0}, which leads to $\gamma_{\rm film}/\gamma_{\rm in}=\mathfrak{F}\,(\Rp/\epsilon_1)\ee^{-4\pi z_0/\lamp}$ for $\pb_0$ oriented parallel to the film ({\it i.e.}, with the rotation axis of the prolate ellipsoid normal to the film and polarization along the diameter), where we have defined the prefactor
\begin{align}
\mathfrak{F}= \frac{8 \pi^4 p_0^2}{\hbar \gamma_{in} \lamp^3}=\frac{\pi^3\epsilon_1^2(1-\varepsilon)^2}{(\epsilon_{\rm b}-\epsilon_1\varepsilon)^2}
\frac{\omega_0}{\gamma_{\rm in}}\,\frac{V}{\lamp^3}, \label{mathfrakFF}
\end{align}
while the rest of the factors ($(\Rp/\epsilon_1)\ee^{-4\pi z_0/\lamp}$) have unity order for $z_0\ll\lamp$ and we have approximated $\omega_0\approx\tilde\omega_0$. Applying this expression to gold oblate ellipsoids of 100\,nm diameter and 20\,nm height ({\it i.e.}, $r=5$), embedded in an $\epsilon_1=2$ medium, and coupled to film polaritons of wavelength $\lamp=200\,$nm, we find $\hbar\omega_0=1.9\,$eV and $\mathfrak{F}=5$ (for silver, we obtain $\hbar\omega_0=2.1$\,eV and $\mathfrak{F}=32$), thus establishing $\gamma_{\rm film}\gg\gamma_{\rm in}$, so that internal inelastic losses play a relatively small role and the maximum possible polarizability derived in the present work can be indeed approached. Numerical results are presented in the Appendix\ \ref{InfluenceofSize} for gold and silver spheres and disks, further supporting this conclusion for a wide range of attainable geometrical parameters.

\section{Correction Factor $\mathfrak{f}$} 
\label{sec:AppI}

When calculating the emission rate of a point dipole oscillating at frequency $\omega$ through the evaluation of the generated Poynting vector, the result varies depending on whether the surrounding $\epsilon_1$ material is either ({\it i}) fully overlapping the dipole or ({\it ii}) just extending outside an empty cavity of vanishingly small size. When the dipole is associated with an excited state of an atom or molecule, model {\it ii} should be used, as the $\epsilon_1$ material does not penetrate inside the interstitial regions of the emitter; for an embedding spherical cavity, direct integration of the outgoing Poynting vector over a surface right outside it reveals an emission rate identical to that of model {\it i}, but multiplied by the so-called spherical-local-field correction factor\cite{YGB1988} $\mathfrak{f}^2$, where $\mathfrak{f}=3/(\epsilon_1+2)$. This correction remains unchanged even in the presence of additional neighboring structures ({\it e.g.}, a film). For the metal nanoparticles under consideration, this correction is automatically included in the effective excitation dipole $\pb_0$ (see Eq.\ \eqref{alphapart}), so model {\it i} should be applied and no correcting factor is needed ({\it i.e.}, $\mathfrak{f}=1$).}

% \vfill\null

\section{Poynting-Vector-Based Calculation of the Scattered Power}
\label{Poynting1}

In this section, we provide an alternative derivation of Eq.\ \eqref{Pscat} of the main text based on the Poynting vector, for which we perform fully electrodynamic calculations because we need the magnetic field that otherwise vanishes in the quasistatic limit. We focus on the system depicted in Figure\ 1b, with a dipole $\pb$ placed right above the $z=0$ planar interface that separates regions of permittivities $\vep_1$ and $\vep_{\rm m}$. For reference, we divide space into the three regions I, II, and III defined by the conditions $z>0$, $-d<z<0$, and $z<-d$, and filled with materials of permittivities $\vep_1$, $\vep_{\rm m}$, and $\vep_2$, respectively.

The electric field produced by a dipole $\pb$ placed at $\rb'=(0,0,z')$ in a homogeneous medium of permittivity $\epsilon_1$ can be written as \cite{paper090}
\begin{align}
\Eb^\ind{homo} &= \frac{1}{\epsilon_1} \sqpar{k_1^2\pb + (\pb\cdot\vect{\nabla})\vect{\nabla}} \frac{\ee^{\ii k_1 |\rb-\rb'|}}{|\rb-\rb'|} \nonumber\\&=\! \dfrac{\ii}{2\pi\epsilon_1}\!\! \int\!\! \dfrac{d^2{\kparb}}{k_{z1}} \sqpar{k_1^2\pb\! -\! (\pb\cdot\kb^\pm_1)\,\kb^\pm_1}\! \ee^{\ii(\kparb\cdot \Rb + k_{z1} |z-z'|)},
\label{eqS1}
\end{align}
where $\Rb=x\xx+y\yy$, $k=\omega/c$, $k_1=k\sqrt{\epsilon_1}$, $k_{z1}=(k_1^2 - \kpar^2)^{1/2}$, $\kparb=k_x\xx+k_y\yy$, $\kb^\pm_1=\kparb\pm k_{z1}\zz$, the $+$ and $-$ sign must be chosen for $z>z'$ and $z<z'$, respectively, and the square roots are taken to yield a positive imaginary part. Noticing that the unit vector $\hat{\kb}^\pm_1\equiv\kb^\pm_1/k_1$ together with the s and p polarization vectors $\eh_{\rm s} = (-k_y \vers{x} + k_x \vers{y})/\kpar$ and $\eh_{\rm p1}^{\pm} = (\pm k_{z1} \kparb - \kpar^2 \vers{z})/(k_1 \kpar)$ form a complete set of orthonormal vectors, so that $\hat{\kb}^\pm_1 \otimes \hat{\kb}^\pm_1 + \eh_{\rm s} \otimes \eh_{\rm s} + \eh_{\rm p1}^{\pm} \otimes \eh_{\rm p1}^{\pm}$ is the $3\times3$ identity matrix, we can rewrite Eq.\ \eqref{eqS1} as
\begin{align}
\Eb^\ind{homo} = \dfrac{\ii k^2}{2 \pi} \int \dfrac{d^2{\kparb}}{k_{z1}} \sqpar{  (\pb\cdot\eh_{\rm s}) \, \eh_{\rm s} + (\pb\cdot\eh_{\rm p1}^{\pm}) \, \eh_{\rm p1}^{\pm}} \cdot \nonumber \\ \cdot \ee^{\ii(\kparb\cdot \Rb + k_{z1} |z-z'|)}.
\nonumber
\end{align}
From this expression, the magnetic field can be obtained via Ampere's law ({\it i.e.}, $\Hb=\nabla\times\Eb/(\ii k)$), which by making the replacement $\nabla\rightarrow\ii\kb^\pm_1$ inside the integral and using the identities $\hat{\kb}^\pm_1\times\eh_{\rm s}=-\eh_{\rm p1}^{\pm}$ and $\hat{\kb}^\pm_1\times\eh_{\rm p1}^{\pm}=\eh_{\rm s}$, leads to
\begin{align}
\Hb^\ind{homo} = \dfrac{\ii k^2\sqrt{\epsilon_1}}{2 \pi} \int \dfrac{d^2{\kparb}}{k_{z1}} \sqpar{-(\pb\cdot\eh_{\rm s}) \, \eh^\pm_{\rm p1} + (\pb\cdot\eh_{\rm p1}^{\pm}) \, \eh_{\rm s}} \cdot \nonumber \\ \cdot \ee^{\ii(\kparb\cdot \Rb + k_{z1} |z-z'|)}.
\nonumber
\end{align}
As we are interested in obtained the reflected field produced by the dipole in region I, we consider waves moving downward from the dipole position at $\rb'$  ({\it i.e.}, $\eh^-$ vectors), multiply them by the corresponding reflection coefficients $r_{\rm s}$ and $r_{\rm p}$ for s and p polarization, and reverse their orientation toward the upward direction ({\it i.e.}, $\eh^+$). Further taking the dipole position $z'\rightarrow0^+$ immediately above the interface, this procedure leads to the reflected fields
\begin{widetext}
\begin{subequations}
\begin{align}
\Eb^\ind{ref}_{\rm I} &= \dfrac{\ii k^2}{2 \pi} \int \dfrac{d^2{\kparb}}{k_{z1}} \sqpar{\rs\,(\pb\cdot\eh_{\rm s}) \, \eh_{\rm s} + \rp\,(\pb\cdot\eh_{\rm p1}^-) \, \eh_{\rm p1}^+}  \ee^{\ii(\kparb\cdot \Rb + k_{z1}z)}, % \cdot \nonumber \\ \cdot
\label{eq:EI:begining} \\
\Hb^\ind{ref}_{\rm I} &= \dfrac{\ii k^2\sqrt{\epsilon_1}}{2 \pi} \int \dfrac{d^2{\kparb}}{k_{z1}} \sqpar{-\rs\,(\pb\cdot\eh_{\rm s}) \, \eh^+_{\rm p1} + \rp\,(\pb\cdot\eh_{\rm p1}^-) \, \eh_{\rm s}} \ee^{\ii(\kparb\cdot \Rb + k_{z1}z)}. % \cdot \nonumber \\ \cdot 
\label{eq:HI:begining}
\end{align}
\label{eq:beginning}
\end{subequations}
\end{widetext}

\begin{table}
\begin{center}
\begin{tabular}{|c|c|} \hline
{\color{blue} $P(\kparb)$} & {\color{blue} $\int d\varphi_{\kparb}\,P(\kparb)\,\exp(\ii\kparb\cdot\Rb)$} \\ \hline
$1$ & $2 \pi J_0$ \\
$k_x$ &  $2 \ii \pi \kpar \cos \varphi J_1$ \\
$k_y$ &  $2 \ii \pi \kpar \sin \varphi J_1$ \\
$k_x^2$ & $ \pi \kpar^2 (\cos^2 \varphi (-J_2+J_0)
          + \sin^2 \varphi (J_2 + J_0 ))$ \\
$k_x k_y $ & $ -2\pi \kpar^2 \cos \varphi \sin \varphi  J_2$ \\
$k_y^2$ & $ \pi \kpar^2 (\cos^2 \varphi (J_2+J_0)
          + \sin^2 \varphi (-J_2 + J_0 ))$ \\ \hline
\end{tabular}
\end{center}
\caption{Azimuthal integrals needed to evaluate Eq.\ \eqref{eq:beginning} for different polynomials $P(\kparb)$. We use azimuthal coordinates $\kparb=(\kpar,\varphi_{\kparb})$ and $\Rb=(R,\varphi)$, as well as the abbreviation $J_m \equiv J_m(\kpar R)$. Reproduced from Ref.\ \cite{paper053}.}
\label{tab:Bessel}
\end{table}

We aim to study surface-polariton modes supported by the film in region II, and specifically, we concentrate on modes signaled by the poles of $\rp$, which include plasmons in thin films. Incidentally, a similar study could be carried out for the poles of $\rs$, which describe for example some of the guided modes in sufficiently-thick or sufficiently-high-refractive-index planar waveguides. Consequently, we dismiss $\rs$ terms in the present study and approximate $r_{\rm p} \approx \mathcal{R}_{\rm p} \kp/(\kpar-\kp)$ (Eq.\ \eqref{rp} in the main text), where $\kp$ is the in-plane wave vector of the mode under consideration. We first proceed by carrying out the integral over the azimuthal angle of $\kparb$ with the help of the expressions compiled in Table\ \ref{tab:Bessel}. This integration generates Bessel functions $J_n(\kpar R)$, which have the asymptotic behavior $J_n(\kpar R)\approx(2\pi\kpar R)^{-1/2} [\ee^{\ii (\kpar R- \pi/4 - n \pi/2)} + \ee^{-\ii (\kpar R- \pi/4 - n \pi/2)}]$ in the $|\kpar R|\gg1$ limit. For the remaining radial integral, we proceed in a similar way as explained in the above Appendix sections: the contribution to the field associated with the mode at $\kpar=\kp$ is dominated by the pole in $\rp$; we then extend the integral down to $\kpar=-\infty$ and integrate in the complex $\kpar$ plane by closing contours in the upper and lower half-planes for the terms $\propto\ee^{\ii \kpar R}$ and $\propto\ee^{-\ii \kpar R}$, respectively; finally, noticing that $\mathrm{Im}\{k_{\rm p}\}>0$, only the first of these terms is found to make a contribution to the resulting scattered field, which becomes
\begin{widetext}
\begin{subequations}
\begin{align}
\Eb^\ind{scat}_{\rm I}\!\! &\approx \!\! \sqrt{2 \pi} \ee^{-\ii \pi/4} \dfrac{\mathcal{R}_{\rm p}}{\vep_1}\! \frac{\ee^{k_\ind{p}(\ii R - z/\Omega)}}{\sqrt{k_\ind{p} R}}\!  \frac{k_\ind{p}^3}{\Omega} \ccpar{ \ii \pb_{\parallel}\! \cdot\! \vers{R}\! +\! \Omega p_z\! } (\vers{R}\! +\! \ii \Omega \vers{z}),
\label{eq:E:I} \\
\Hb^\ind{scat}_{\rm I}\!\! &\approx \!\! \sqrt{2 \pi}  \ee^{-3\ii \pi/4}  \mathcal{R}_{\rm p} \frac{\ee^{k_\ind{p}(\ii R - z/\Omega)}}{\sqrt{k_\ind{p} R}} k k_\ind{p}^2 \ccpar{ \ii \pb_{\parallel}\! \cdot\! \vers{R}\! +\! \Omega p_z\!} \vers{\varphi},
\label{eq:H:I}
\end{align}
\end{subequations}
\end{widetext}
where $\Omega=\ii\kp/\sqrt{k_1^2-\kp^2}$ fully captures the effect of retardation. Indeed, one has $\Omega=1$ in the quasistatic limit, and in particular, Eq.\ \eqref{eq:E:I} then reduces to Eq.\ \eqref{Eprad} of the main text. In what follows we adopt this limit, but still retain an overall factor of $k$ in the expressions for the magnetic field (Eq.\ \eqref{eq:H:I}).

We now proceed in a similar way to obtain the fields in regions II and III, just be replacing the polarization vectors with those propagating in the corresponding media, and by making use of the self-consistent transmission and reflection coefficients of the involved interfaces (see Section\ \ref{sec:ape:Fresnel}). After some lengthy but straightforward algebra, the resulting scattered fields associated with the mode under consideration reduce to
\begin{widetext}
\begin{align}
\Eb^\ind{scat}_{\rm II} &= \sqrt{2 \pi} \ee^{-\ii \pi/4} \dfrac{1}{\sqrt{\vep_1 \vep_\ind{m}}} \frac{\ee^{\ii k_\ind{p} R}}{\sqrt{k_\ind{p} R}} k_\ind{p}^3 \ccpar{ \ii \pb_{\parallel} \cdot \vers{R} + p_z } \clpar{  \mathcal{B}^+_{\rm p} \ee^{-k_\ind{p} (z+d)} (\vers{R} + \ii \vers{z}) - \mathcal{B}^-_{\rm p} \ee^{k_\ind{p} z} (\vers{R} - \ii \vers{z}) },
\nonumber \\
\Hb^\ind{scat}_{\rm II} &= - \ii \sqrt{2 \pi}  \ee^{-\ii \pi/4} \sqrt{\dfrac{\epsilon_\ind{m}}{\epsilon_1}}  \frac{\ee^{\ii k_\ind{p} R }}{\sqrt{k_\ind{p} R}} k k_\ind{p}^2 \sqpar{ \mathcal{B}^+_{\rm p} \ee^{-k_\ind{p} (z+d)} + \mathcal{B}^-_{\rm p} \ee^{k_\ind{p} z}} \ccpar{ \ii \pb_{\parallel} \cdot \vers{R} + p_z } \vers{\varphi},
\nonumber \\
\Eb^\ind{scat}_{\rm III} &= - \sqrt{2 \pi} \ee^{-\ii \pi/4} \dfrac{\mathcal{T}_{\rm p}}{\sqrt{\vep_1 \vep_2}} \frac{\ee^{k_\ind{p}(\ii R + z + d)}}{\sqrt{k_\ind{p} R}}  k_\ind{p}^3 \ccpar{ \ii \pb_{\parallel} \cdot \vers{R} + p_z } (\vers{R} - \ii \vers{z}),
\nonumber \\
\Hb^\ind{scat}_{\rm III} &= - \ii \sqrt{2 \pi}  \ee^{-\ii \pi/4} \sqrt{\dfrac{\epsilon_2}{\epsilon_1}} \mathcal{T}_{\rm p} \frac{\ee^{k_\ind{p}(\ii R + z + d)}}{\sqrt{k_\ind{p} R}} k k_\ind{p}^2 \ccpar{ \ii \pb_{\parallel} \cdot \vers{R} + p_z } \vers{\varphi},
\nonumber
\end{align}
\end{widetext}
where the dimensionless coefficients $\mathcal{T}_{\rm p}$ and $\mathcal{B}^{\pm}_{\rm p}$ are implicitly defined as the residues in $t_{\rm p} \approx \mathcal{T}_{\rm p} \kp/(\kpar-\kp)$ and $\beta^{\pm}_{\rm p} \approx \mathcal{B}^{\pm}_{\rm p} \kp/(\kpar-\kp)$, which are good approximations to the coefficients $t_{\rm p}$ and $\beta^{\pm}_{\rm p}$ (Eqs.\ \eqref{eq:Fresnel:rt} and \eqref{eq:Fresnel:beta} in Section\ \ref{sec:ape:Fresnel}) near the mode pole $\kpar=\kp$.

We are now prepared to calculate the time-averaged Poynting vector $\left\langle \Sb^{\rm scat} \right\rangle = [c/(2\pi)] \mathrm{Re}\left\{\Eb^{\rm scat} \times (\Hb^{\rm scat})^{\ast}\right\}$. Assuming that $k_\ind{p}$ is approximately real, the radial component is found to be
\begin{align}
\left\langle \Sb^{\rm scat}_{\rm I}\cdot\RR \right\rangle \! &= \! \dfrac{\omega k_\ind{p}^4}{\epsilon_1 R} |\mathcal{R}_{\rm p}|^2 \ee^{-2 k_\ind{p} z} \ccpar{ \ii \pb_{\parallel}\! \cdot\! \vers{R}\! +\! p_z\!}, \nonumber\\
\left\langle \Sb^{\rm scat}_{\rm II}\cdot\RR \right\rangle \! &= \! \dfrac{\omega k_\ind{p}^4}{\epsilon_1 R} \sqabs{\mathcal{B}_{\rm p}^+ \ee^{-k_\ind{p} (z+d)} + \mathcal{B}_{\rm p}^- \ee^{k_\ind{p} z}} \ccpar{ \ii \pb_{\parallel}\! \cdot\! \vers{R}\! +\! p_z\!}, \nonumber\\
\left\langle \Sb^{\rm scat}_{\rm III}\cdot\RR \right\rangle \! &= \! \dfrac{\omega k_\ind{p}^4}{\epsilon_1 R} |\mathcal{T}_{\rm p}|^2 \ee^{2 k_\ind{p} (z+d)} \ccpar{ \ii \pb_{\parallel}\! \cdot\! \vers{R}\! +\! p_z\!}. \nonumber
\end{align}
It should be noted that, although the magnetic field is proportional to $k$ and thus vanishes in the quasistatic limit, the Poynting vector introduces a factor of $c$, rendering a finite product $kc=\omega$. Finally, we obtain the power scattered by the dipole by integrating over the surface of a cylinder or large radius $R$ centered at the dipole and oriented perpendicularly to the film:
\begin{align}
P^\ind{scat} &= \int_{-\infty}^{\infty} dz\ \int_{0}^{2\pi} R\,d\varphi\, \left\langle \Sb^{\rm scat} \cdot \vers{R} \right\rangle \nonumber \\ &= \pi \omega k_\ind{p}^3 \dfrac{\Lambda}{\vep_1} \ccpar{ |\pb_{\parallel}|^2/2 + |p_z|^2 },
\nonumber
\end{align}
where $\Lambda= |\mathcal{R}_{\rm p}|^2 + |\mathcal{T}_{\rm p}|^2 + 2 \ee^{-k_\ind{p} d}[ k_\ind{p} d (\mathcal{B}_{\rm p}^+ (\mathcal{B}_{\rm p}^-)^{\ast} + \mathcal{B}_{\rm p}^- (\mathcal{B}_{\rm p}^+)^{\ast}) + \sinh(k_\ind{p} d) (|\mathcal{B}_{\rm p}^+|^2 + |\mathcal{B}_{\rm p}^-|^2) ]$. When explicitly working out $\mathcal{R}_{\rm p}$, $\mathcal{T}_{\rm p}$, $\mathcal{B}_{\rm p}^{\pm}$ (Eq.\ \eqref{eq:Fresnellast}) and $k_\ind{p}$ (Eq.\ \eqref{lampd} in the main text), this expression reduces to $\Lambda=2|\mathcal{R}_{\rm p}|$, thus recovering Eq.\ \eqref{Pscat} of the main text for the power scattered by the dipole into surface polaritons.
 
\section{Fresnel Coefficients}
\label{sec:ape:Fresnel}

The calculations presented above involve the Fresnel coefficients of the film, which one can calculate in a Fabry-Perot model from the reflection and transmission coefficients $r_{\nu,ij}$ and $t_{\nu,ij}$ for polarization $\nu=$s,p and incidence from each medium $i$ on the interface with each of its surrounding media $j$. We consider homogeneous, isotropic media $i=1$, m, and 2 in regions I, II, and III, respectively, where the central (film) region has thickness $d$. The reflection and transmission coefficients of the film for incidence from the top medium 1 then reduce to
\begin{subequations}
\begin{align}
r_{\nu}&=r_{\nu,1{\rm m}}+\frac{t_{\nu,1{\rm m}} r_{\nu,{\rm m}2} t_{\nu,{\rm m}1} \ee^{2 \ii k_{z{\rm m}}d}}{1-r_{\nu,{\rm m}2}r_{\nu,{\rm m}1}\ee^{2\ii k_{z{\rm m}}d}},\\
t_{\nu}&=\frac{t_{\nu,1{\rm m}} t_{\nu,{\rm m}2} \ee^{\ii k_{z{\rm m}}d}}{1-r_{\nu,{\rm m}2}r_{\nu,{\rm m}1}\ee^{2 \ii k_{z{\rm m}}d}},
\label{eq:Fresnel:rt}
\end{align}
\end{subequations}
where $k_{z{\rm m}}$ is the light wave vector in medium $m$. In the central region m, the field can be described as a superposition of waves propagating upward and downward:
\begin{subequations}
\begin{align}
\beta^+_{\nu}&=\frac{t_{\rm \nu, 1m} r_{\nu,{\rm m}2} \ee^{\ii k_{z{\rm m}} d}}{ 1 - r_{\nu,{\rm m}2}r_{\nu,{\rm m}1} \ee^{2 \ii k_{z{\rm m}} d}}, \\
\beta^-_{\nu}&=\frac{t_{\nu,1{\rm m}}}{ 1 - r_{\nu,{\rm m}2}r_{\nu,{\rm m}1} \ee^{2 \ii k_{z{\rm m}} d}}.
\end{align}
\label{eq:Fresnel:beta}
\end{subequations}
Near a polariton mode of in-plane wave vector $\kp$, these coefficients exhibit a divergence that we isolate to obtain the fields scattered by a dipole (Section\ \ref{Poynting1}). More precisely, for the structure depicted in Figure\ 1b of the main text, we find
\begin{align}
r_{\rm p} \equiv \mathcal{R}_{\rm p} \dfrac{k_\ind{p}}{\kpar-k_\ind{p}}, \nonumber\\
\beta^{\pm}_{\rm p} \equiv \mathcal{B}_{\rm p}^{\pm} \dfrac{k_\ind{p}}{\kpar-k_\ind{p}}, \nonumber\\
t_{\rm p} \equiv \mathcal{T}_{\rm p} \dfrac{k_\ind{p}}{\kpar-k_\ind{p}}, \nonumber
\end{align}
where
\begin{subequations}
\begin{align}
\mathcal{R}_{\rm p} &= \dfrac{1}{k_\ind{p} d} \ccpar{ \dfrac{2 \vep_1 \vep_\ind{m}}{\vep_1^2 - \vep_{\rm m}^2} }, \\ 
\mathcal{T}_{\rm p} &= \dfrac{1}{k_\ind{p} d} \sqpar{ \dfrac{2  \sqrt{ \vep_1 \vep_\ind{m} } \sqrt{ \vep_\ind{m} \vep_2 } }{(\vep_1 + \vep_\ind{m})(\vep_\ind{m} + \vep_2)} } \ee^{-k_\ind{p} d}, \\
\mathcal{B}_{\rm p}^- &= \dfrac{1}{k_\ind{p} d} \ccpar{ \dfrac{\sqrt{ \vep_1 \vep_\ind{m} }}{\vep_1 + \vep_\ind{m}} }, \\
\mathcal{B}_{\rm p}^+ &= \dfrac{1}{k_\ind{p} d} \ccpar{ \dfrac{\sqrt{ \vep_1 \vep_\ind{m} }}{\vep_1 - \vep_\ind{m}} } \ee^{k_\ind{p} d}.
\end{align}
\label{eq:Fresnellast}
\end{subequations}
 
\begin{figure*}[htbp]
\centering
\includegraphics[width=0.99\textwidth]{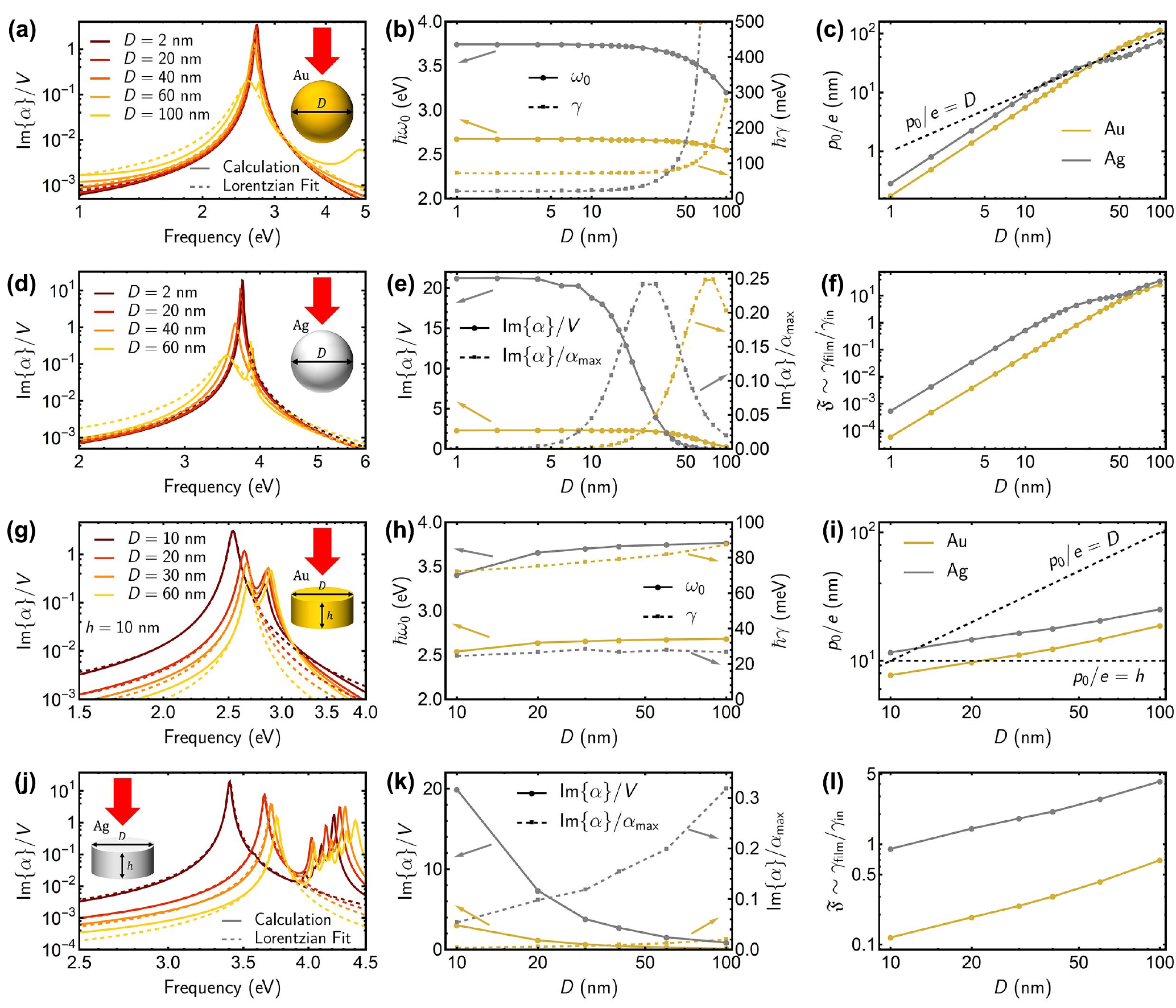}
\caption{Size-dependence analysis of the polarizability of metallic spheres \textbf{(a-f)} and 10-nm-high disks \textbf{(g-l)}. \textbf{(a,d,g,j)} Spectral variation of the polarizability (solid curves) for particles made of (a,g) gold and (d,j) silver, along with Lorentzian fits (Eq.\ \eqref{alphaSI}, dashed curves) of the lowest-energy resonance ($\omega = \omega_0$). \textbf{(b,h)} Variation of the resonance position $\omega_0$ (left axis, solid curves) and width $\gamma$ (right axis, dashed curves) extracted from the Lorentzian fits in (a,d,g,j). \textbf{(e,k)} Resonant value of the polarizability (${\rm Im}\{\alpha(\omega_0)\}$) as a function of particle diameter $D$. We normalize the polarizability either to the volume of the particles (left axis, solid curves) or to the maximum polarizability of a lossless two-level scatterer $\alpha_{\rm max}=3\lambda_0^3/(16\pi^3\sqrt{\epsilon_1})$ (right axis, dashed curves). \textbf{(c,i)} Effective dipole moment $p_0$ extracted from the Lorentzian fits. \textbf{(f,l)} Dimensionless prefactor $\mathfrak{F}$ defined in Eq.\ \eqref{mathfrakFF}, assuming a polariton wavelength $\lamp=200\mbox{ nm}$. The metals are described using a Drude-like model $\epsilon_{\rm m}=\epsilon_{\rm b}-\omega_{\rm bulk}^2/[\omega(\omega+\ii\gamma)]$ with parameters $\epsilon_{\rm b}=9.5$, $\hbar\omega_{\rm bulk}=9.06$\,eV, and $\hbar\gamma=71\,$meV for gold; and $\epsilon_{\rm b}=4.0$, $\hbar\omega_{\rm bulk}=9.17$\,eV, and $\hbar\gamma=21\,$meV for silver.}
\label{fig:Size}
\end{figure*}

\section{Influence of Size and Shape of the Scatterer}
\label{InfluenceofSize}

In the main text, we consider point-like scatterers. In this section, we discuss the validity of this approximation when the scatterers are taken to be metallic particles of finite size. In particular, we study spheres and disks made of either gold or silver, which we simulate numerically using the boundary element method (BEM) \cite{paper040}. A summary of the results is presented in Figure\ \ref{fig:Size}. In particular, Figure\ \ref{fig:Size}(a,d,g,j) show the imaginary part of the effective polarizability normalized to the particle volume (solid curves), as obtained by dividing the extinction cross-section by $8\pi^2/(\lambda_0\sqrt{\epsilon_1})$. We now concentrate on the lowest-energy plasmon feature of each spectrum and analyze it by approximating the polarizability as
\begin{align}
\alpha = \dfrac{p_0^2/\hbar}{\omega_0 - \omega - \ii \gamma/2}
\label{alphaSI}
\end{align} 
(similar to Eq.\ \eqref{alphafull} in the main text), where $p_0$, $\omega_0$, and $\gamma$ are fitting parameters. This leads to the Lorentzian profiles represented in Figure\ \ref{fig:Size}(a,d,g,j)  as dashed curves. Given the relatively small size of the particles compared with the resonance light wavelength, retardation does not play a major role, so plasmons show a nearly size-independent frequency and width, although the latter exhibits a rapid increase for silver (gold) spheres of diameter $D$ above $\sim30\,$nm (50\,nm) due to radiative losses. Remarkably, we find an optimum size for the spheres, near the onset of radiative losses, for which the peak polarizability reaches $\sim25$\% of the maximum possible value for a dipolar scatterer (Figure\ \ref{fig:Size}e, see caption). Additionally, for silver disks the peak polarizability reaches $30$\% of the maximum possible value  (Figure\ \ref{fig:Size}k). We are ultimately interested in the parameter $\mathfrak{F}$ (see Appendix\ \ref{sec:AppH}), which reaches values $\sim35$ for silver spheres of 100\,nm diameter coupling to 200\,nm plasmons, and slightly lower values for gold spheres (Figure\ \ref{fig:Size}f). In contrast, disks exhibit much lower values of $\mathfrak{F}$, in consonance with their smaller volume.

\section{Results with Retardation}
\label{sec:Retardation}

\begin{figure*}[htb]
\centering
\includegraphics[width=0.87\textwidth]{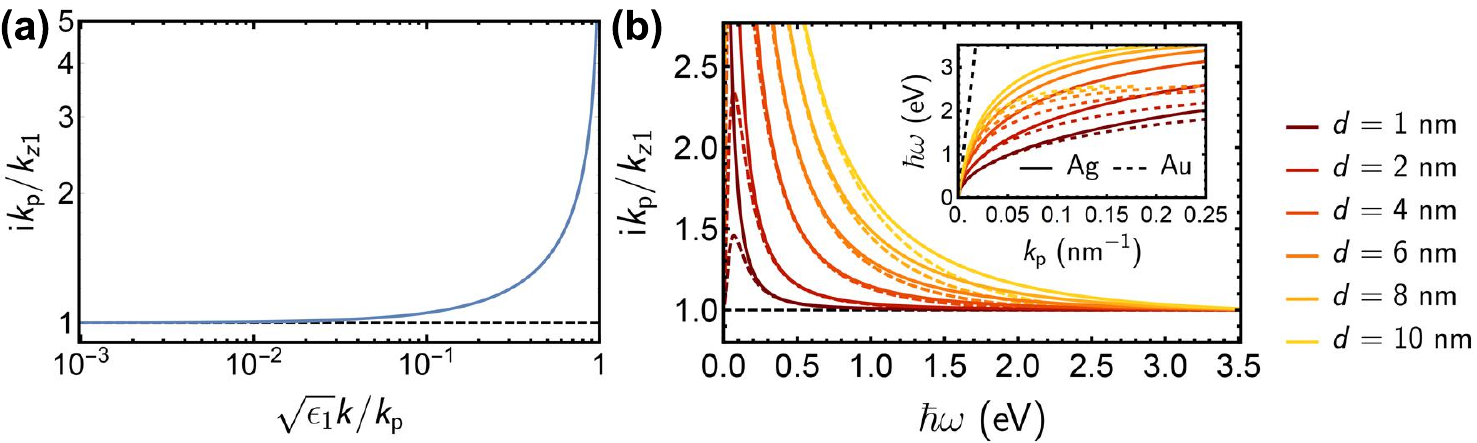}
\caption{\textbf{(a)} Retardation factor $\Omega = \ii \kp/\kz$ as a function of the ratio $\sqrt{\epsilon_1}k/\kp=\sqrt{\epsilon_1}\lamp/\lambda_0$. \textbf{(b)} Retardation factor for the lowest-energy plasmons sustained by gold (solid curves) and silver (dashed curves) films of different thicknesses $d$ embedded in an $\epsilon=2$ material. The inset shows the corresponding plasmon dispersion relations. The metals are described using the same dielectric functions as in Figure\ \ref{fig:Size}.}
\label{fig:Gamma}
\end{figure*}

The results presented in the main text are obtained within the quasistatic limit ($c \to \infty$). A straightforward extension of the derivations presented in the previous sections shows that the main results of the paper can be easily corrected to include retardation by just using a single correction factor
\begin{align}
\Omega \equiv \dfrac{\ii \kp}{k_{z1}}
\nonumber
\end{align}
according to the the summary presented in Table\ \ref{tab:ED}. Specifically, the result $\sigma_{\rm max} \propto 2 \lambda_\ind{p}/\pi$ is maintained, independent of the dielectric and geometrical properties of the film even when retardation is included. Likewise, we also recover the results
\begin{align}
\dfrac{\sigma^{\rm ext}_{\rm point}}{\sigma_{\rm max}} &= \dfrac{\alpha_i}{\mu_i}, \nonumber \\
\dfrac{\sigma^{\rm scat}_{\rm point}}{\sigma_{\rm max}} &= \left(\dfrac{\alpha_i}{\mu_i}\right)^2, \nonumber
\end{align}
indicating that Figure\ 2b of the main text is a universal result. A plot of $\Omega$ as a function of $\sqrt{\epsilon_1} k/\kp$ (Figure\ \ref{fig:Gamma}a) shows that the quasistatic limit is an excellent approximation as long as $k\lesssim0.1\kp$. This is for example the case in 4\,nm silver and gold films above 1.5\,eV plasmon energy (Figure\ \ref{fig:Gamma}b).

\begin{table*}
\begin{center}
\begin{tabular}{|c|c|c|c|}
\cline{3-4}
\multicolumn{2}{c|}{\multirow{2}{*}{}} & {\color{blue} \textbf{Quasistatic Limit}} & {\color{blue} \textbf{Including Retardation}}\\\hline
%\endhead -------------------------------------------------
\parbox[c]{4mm}{\multirow{21}{*}{\rotatebox[origin=c]{90}{\color{blue} {\bf Point Scatterer}}}}
% Electric Field Shape
& $\mathbf{E}^{\rm scat}_{\rm point}$ &
$ E^{\rm scat}_{\hat{\bf R}} \ccpar{ \hat{\bf R} + \ii \hat{\bf z} }\dfrac{\ee^{\kp(\ii R - z)}}{\sqrt{\kp R}}$ &
$E^{\rm scat}_{\hat{\bf R}} \ccpar{ \hat{\bf R} + \ii \Omegared \hat{\bf z}}\dfrac{\ee^{\kp ( \ii R - z{\color{red}/\Omegared)}}}{\sqrt{\kp R}}$
\\ % Electric Field Detail
& $E^{\rm scat}_{\hat{\bf R}}$ &
$\sqrt{2 \pi} \ee^{-\ii \pi/4} \dfrac{\Rp}{\epsilon_1} \kp^3 \ccpar{ \ii \mathbf{p}_{\parallel} \cdot \hat{\mathbf{R}} + p_z}$
& $\sqrt{2 \pi} \ee^{-\ii \pi/4} \dfrac{\Rp}{\epsilon_1} \kp^3 \ccpar{ \ii \mathbf{p}_{\parallel} \cdot \hat{\mathbf{R}}{\color{red} /\Omegared} + p_z } $
\\ % Coupling Cross-section
& $\sigma^{\rm in-coup}_{\rm point}$ &
$\dfrac{|\Rp|}{\epsilon_1^{3/2}} \dfrac{(2\pi)^6}{\lamp^3 \lambda_0}  \sqpar{ A_-^\theta \dfrac{|\alpha_x|^2}{2} + A_+^\theta |\alpha_z|^2}$ &
$ \dfrac{|\Rp|}{\epsilon_1^{3/2}} \dfrac{(2\pi)^6}{\lamp^3 \lambda_0} \sqpar{ A_-^\theta \dfrac{|\alpha_x|^2}{2|\Omegared|} +  A_+^\theta {\color{red} |\Omegared|} |\alpha_z|^2 }  $
\\ % Scattering Cross-section
&
$\sigma^{\rm scat}_{\rm point}$
&
$ \dfrac{|\Rp|^2}{\epsilon_1^2} \dfrac{(2\pi)^7}{ \lamp^5 } \sqpar{ |\alpha_x|^2 + |\alpha_z|^2 }$
&
$\dfrac{|\Rp|^2}{\epsilon_1^2} \dfrac{(2\pi)^7}{\lamp^5 }  \sqpar{ |\alpha_x|^2{\color{red} /|\Omegared|^2} + |\alpha_z|^2 }$
\\ % Extinction Cross-section
& $\sigma^{\rm ext}_{\rm point}$ &
$\dfrac{|\Rp|}{\epsilon_1} \dfrac{16 \pi^3}{\lamp^2} \Imm \clpar{ \alpha_x + \alpha_z}$ &
$\dfrac{|\Rp|}{\epsilon_1} \dfrac{16 \pi^3}{ \lamp^2} \Imm \clpar{ \alpha_x{\color{red}/\Omegared} + \alpha_z}$
\\ % \mu_x
& $\mu_x$ &
$\dfrac{\epsilon_1 \lamp^3}{4 \pi^4 |\Rp|}$ &
${\color{red} |\Omegared|} \dfrac{\epsilon_1 \lamp^3 }{4 \pi^4 |\Rp|}$
\\ % \mu_z
& $\mu_z$ &
$\dfrac{\epsilon_1 \lamp^3}{8 \pi^4 |\Rp|}$ &
$\dfrac{\epsilon_1 \lamp^3}{8 \pi^4 |\Rp|}$
\\ % max Extinction, Scattering
& $\sigma^{\rm max}$ &
$\dfrac{2 \lamp}{\pi} \times \begin{cases} 2, & x \\ 1, & z \end{cases}$ &
$\dfrac{2 \lamp}{\pi} \times \begin{cases} 2, & x \\ 1, & z \end{cases}$
\\ % max Coupling Cross-section
& $\max\clpar{\sigma^{\rm in-coup}_{\rm point}}$ &
$\dfrac{2 \sqrt{\epsilon_1}}{\pi^2 |\Rp|} \dfrac{\lamp^3}{\lambda_0}\times \begin{cases} A_-, & \mbox{p-pol},\theta=0 \\ A_0, & \mbox{s-pol} \end{cases}$ &
${\color{red} |\Omegared|} \dfrac{2 \sqrt{\epsilon_1}}{\pi^2 |\Rp|} \dfrac{\lamp^3}{\lambda_0}\times
\begin{cases} A_-, & \mbox{p-pol},\theta=0 \\ A_0, & \mbox{s-pol} \end{cases}$
\\ % Npol/Nphot
& $\left.\dfrac{N_{\rm polariton}}{N_{\rm photon}}\right|_{\rm point}$ &
$\dfrac{4 g \epsilon_1^{3/2}}{\pi |\Rp|} \left( \dfrac{\lamp}{\lambda_0} \right)^3 \times \begin{cases} A_-, & \mbox{p-pol},\theta=0 \\ A_0, & \mbox{s-pol} \end{cases}$ &
${\color{red} |\Omegared|} \dfrac{4 g \epsilon_1^{3/2}}{\pi |\Rp|}  \left( \dfrac{\lamp}{\lambda_0} \right)^3 \times \begin{cases} A_-, & \mbox{p-pol},\theta=0 \\ A_0, & \mbox{s-pol} \end{cases}$
\\\hline %------------------------------------------------------
\parbox[c]{4mm}{\multirow{16}{*}{\rotatebox[origin=c]{90}{\color{blue} {\bf Line Scatterer}}}}
& $\mathbf{E}^{\rm scat}_{\rm line}$ &
$ E^{\rm scat}_{\hat{\bf R}} \ccpar{ \hat{\bf x} + \ii \hat{\bf z} }\ee^{\kp(\ii x - z)}$ & 
$E^{\rm scat}_{\hat{\bf R}} \ccpar{ \hat{\bf x} + \ii \Omegared \hat{\bf z}}{\ee^{\kp ( \ii x - z{\color{red}/\Omegared)}}}$
\\ % Electric Field Detail, Line
& $E^{\rm scat}_{\hat{\bf R}}$ &
$2 \pi \kp^2 \dfrac{\Rp}{\epsilon_1} \ccpar{ \ii \mathbf{\Pl}_x + \Pl_z}$ &
$2 \pi \kp^2 \dfrac{\Rp}{\epsilon_1} \ccpar{ \ii \mathbf{\Pl}_x{\color{red} /\Omegared} + \Pl_z  } $
\\ % Coupling Cross-section, line
& $\sigma^{\rm in-coup}_{\rm line}$ &
$\dfrac{|\Rp|}{\epsilon_1^{3/2}}\dfrac{2(2\pi)^5}{\lamp^2 \lambda_0} \sqabs{ \Al_x + \Al_z }$ &
${\color{red} \dfrac{1}{|\Omegared|}} \dfrac{|\Rp|}{\epsilon_1^{3/2}}\dfrac{2(2\pi)^5}{\lamp^2 \lambda_0} \sqabs{ \Al_x + \Omegared \Al_z }$
\\ % r
& $r$ &
$\ii \dfrac{\Rp}{\epsilon_1} \dfrac{(2 \pi)^3}{\lamp^2} (\Al_x + \Al_z)$ &
$\ii \dfrac{\Rp}{\epsilon_1 } \dfrac{(2 \pi)^3}{\lamp^2} \ccpar{ \Al_x{\color{red} /\Omegared} + \Al_z }$
\\ % \mu_x
& $\mu_x$ &
$\dfrac{\epsilon_1 \lamp^2}{(2 \pi)^3 |\Rp|}$ &
${\color{red} |\Omegared|} \dfrac{\epsilon_1 \lamp^2  }{(2 \pi)^3 |\Rp|}$
\\ % \mu_z
& $\mu_z$ &
$\dfrac{\epsilon_1 \lamp^2}{(2 \pi)^3 |\Rp|}$ &
$\dfrac{\epsilon_1 \lamp^2}{(2 \pi)^3 |\Rp|}$
\\ % max Coupling Cross-section, line
& $\max\clpar{\sigma^{\rm in-coup}_{\rm line}}$ &
$\dfrac{\sqrt{\epsilon_1}}{\pi |\Rp|} \dfrac{\lamp^2}{\lambda_0}$ &
${\color{red} |\Omegared|} \dfrac{\sqrt{\epsilon_1}}{\pi |\Rp|} \dfrac{\lamp^2}{\lambda_0}$
\\ % Npol/Nphot
& $\left.\dfrac{N_{\rm polariton}}{N_{\rm photon}}\right|_{\rm line}$ &
$\dfrac{g \epsilon_1}{\pi |\Rp|} \left( \dfrac{\lamp}{\lambda_0} \right)^2 $ &
${\color{red} |\Omegared|} \dfrac{g \epsilon_1 }{\pi |\Rp|} \dfrac{\lamp^2}{\lambda_0^2} $
\\\hline
\end{tabular}
\end{center}
\label{tab:ED}
\caption{Summary of results in the main text (quasistatic limit) and their generalization to include retardation (rigorous corrections in red), where we have defined the parameter $\Omega \equiv \ii \kp/k_{z1}$ ($=1$ in the quasistatic limit) with $k_{z1}=\sqrt{\vep_1 \omega^2/c^2 - \kp^2}$. We have also defined $A_-^\theta \equiv A_- \cos^2(\theta)$ and $A_+^\theta \equiv A_+ \sin^2(\theta)$, which are used in $\sigma^{\rm in-coup}_{\rm point}$.}
\end{table*}

\section*{ACKNOWLEDGMENTS}

F.J.G.A. would like to thank D. Basov and M. Fogler for stimulating and fruitful discussions. This work has been supported in part by the Spanish MINECO (MAT2017-88492-R and SEV2015-0522), the ERC (Advanced Grant 789104-eNANO), the Catalan CERCA Program, and Fundaci\'o Privada Cellex. EJCD acknowledges financial support through a "la Caixa" INPhINIT Fellowship Grant (LCF/BQ/DI17/11620057) and EU Marie Sk\l{}odowska-Curie grant (No. 713673).

%\newpage
\clearpage
%\bibliographystyle{apsrev}
%\bibliography{../../../bibtex/refs}

\end{document}